\documentclass[USenglish]{article}	

\usepackage[utf8]{inputenc}				
\usepackage[big,online]{dgruyter}	
\usepackage{lmodern} 
\usepackage{microtype}
\usepackage{placeins}

\theoremstyle{dgthm}

\theoremstyle{dgdef}

\usepackage{csquotes}
\usepackage{mathtools}
\usepackage{bm}
\usepackage{amsfonts, amsmath, amssymb}
\usepackage{siunitx}
\usepackage{graphicx}
\usepackage{booktabs}
\usepackage{tabularx}
\usepackage{multirow}
\usepackage{tikz}
\usepackage{relsize}
\usepackage{setspace}
\usetikzlibrary{positioning, arrows.meta, quotes}
\usetikzlibrary{shapes, calc}
\usetikzlibrary{bayesnet}
\usepackage{soul}

\definecolor{myteal}{HTML}{008080}
\definecolor{myorange}{HTML}{D35400}

\usepackage{algorithm2e}
\SetKwComment{Comment}{/* }{ */}
\RestyleAlgo{ruled} 
\usepackage[
    style=authoryear,       
    backend=bibtex,          
    natbib = true,
    doi=true,               
    url=false,              
    isbn=false,             
    maxnames=2,             
    maxbibnames=99, 
    giveninits=true         
]{biblatex}

\DeclareBibliographyDriver{article}{%
  \printnames{author}%
  \setunit{\addspace}%
  \printtext[parens]{\printfield{year}}%
  \newunit\newblock
  \printfield[title]{title}%
  \newunit\newblock
  \printfield[journaltitle]{journaltitle}
  \setunit{\addcomma\space}%
  \printfield{volume}%
  \iffieldundef{number}{}{%
   \printtext{\mkbibparens{\printfield{number}}}%
  }%
  \setunit{\addcolon\space}%
  \printfield{pages}%
  \newunit\newblock
  \printfield{doi}%
  \finentry
}

\DeclarePairedDelimiter\floor{\lfloor}{\rfloor}

\begin{document}

	\articletype{Research Article}
	\received{Month	DD, YYYY}
	\revised{Month	DD, YYYY}
  \accepted{Month	DD, YYYY}
  \journalname{De~Gruyter~Journal}
  \journalyear{YYYY}
  \journalvolume{XX}
  \journalissue{X}
  \startpage{1}
  \aop
  \DOI{10.1515/sample-YYYY-XXXX}

\title{Bayesian Conway-Maxwell-Poisson model with spike-and-slab priors for dispersed count data with application to football scores}
\runningtitle{A Bayesian CMP model with Spike-and-slab priors}

\author*[1]{Nick Zhang}
\author[2]{Riccardo Rastelli}
\author[2,3]{Nial Friel} 
\runningauthor{N.~Zhang et al.}
\affil[1]{\protect\raggedright 
University College Dublin, Department of Mathematics and Statistics, Dublin, Ireland, e-mail: nick.zhang@ucdconnect.ie}
\affil[2]{\protect\raggedright 
University College Dublin, Department of Mathematics and Statistics, Dublin, Ireland}
\affil[3]{\protect\raggedright 
Insight Centre for Data Analytics, University College Dublin, Dublin, Ireland}
	
	
\abstract{
Statistical modelling for goals scored in football is typically achieved using the Poisson distribution and its variants. Here we propose a Bayesian framework for modelling under- and over-dispersion in count data by combining the Conway-Maxwell-Poisson (CMP) likelihood with a spike-and-slab (SAS) prior on unit-specific dispersion parameters. The proposed methodology generalizes Poisson-based count data models by treating equidispersion as an explicit baseline, and offering probabilistic quantification of departures from this regime, while simultaneously estimating their magnitude. Posterior inference is performed through a tailored Metropolis-within-Gibbs sampler that handles the doubly-intractable likelihood and provides efficient posterior exploration. The new method is examined using simulated data to confirm its ability to capture non-equidispersion, and applied to English Premier League (EPL) data. Dispersion is modeled at the team level and linked to goal-scoring behaviour, and allows for thresholding mechanisms to distinguish teams based on their posterior probability of non-equidispersion. The results reveal heterogeneities in team-specific dispersion in the EPL, and demonstrate improvements in both model fit and predictive performance with respect to the standard Poisson model.}


\maketitle

\section{Introduction}\label{ch:intro_litreview}
Football outcomes modelling is commonly divided into result-based approaches, typically through some categorical distribution, or goal-based approaches, based on discrete distributions for count data \citep{Egidi21}, which will be the focus of the current work. Some of the key assumptions underlying this class of models are the distributional form of the likelihood, the specification of the model's parameters, and the dependence structure of the two scores in a match. 

In the vast majority of the literature, goal-based approaches assume an underlying Poisson distribution, a natural choice whenever modelling low-count discrete events such as football goals. Its usage can be traced back to \citet{moroney1951facts}, who first proposes its application for football score analysis. One of the assumptions of the Poisson likelihood is the \textit{equidispersion} of the observations, whereby the mean and variance are equal. The Negative Binomial distribution, which allows for overdispersion, has been studied in the literature to address this assumption (e.g. \cite{Pollard85}), although the gain in model fit is usually very minor at high levels of aggregation of goals.
Similar conclusions are reached through the Weibull count model of \citet{boshnakov2017bivariate}, which is able to account for both over- and under-dispersion, where no significant dispersion is identified when aggregating over home and away goals. In this paper, we study a more granular aggregation of goals: by examining the frequencies of goals scored by each team throughout a season, we find substantial differences in the overall dispersion in scoring behaviour. To address this, we consider the Conway-Maxwell-Poisson (CMP) distribution \citep{ConwayMaxwell62} which is able to flexibly model over- and under-dispersion, and has been receiving increasing attention in the recent statistical modelling literature. A novelty of our work is that we employ a spike-and-slab (SAS) prior \citep{GeorgeMcCulloch1993} over the dispersion parameters in the CMP distribution, motivated by the heterogeneities between the teams in the overall magnitude and direction of \textit{non-equidispersion}, defined as departures from \textit{equidispersion}. Our specification distinguishes between two regimes of the data-generating process: an equidispersed Poisson distribution and a non-equidispersed CMP distribution. The resulting SAS framework yields interpretable posterior probabilities of each team's membership in either regime, enabling a clear classification through a thresholding rule.

As for the model's parameters, we build upon the seminal work by \citet{Maher1982modelling}, widely recognized as the first one to model the scoring rate of the Poisson likelihood as functions of latent variables representing the offensive and defensive capabilities of the two teams represented in each match. A home advantage is often included in order to account for the higher number of goals scored by the home team. This type of specification provides interpretable parameters, and remains to this day a foundation for countless statistical models in football \citep{Ridall2025,Whitaker2021} and other sports such as basketball \citep{Ruiz2014}. 

Regarding the dependence structure of the two scorelines, \citet{Maher1982modelling} employs two \textit{conditionally independent} Poisson distributions for the two scores in the match, with dependence arising indirectly through the team-specific parameters. \citet{Dixon1997} extend this framework with a focus on its predictive capabilities, and introduce a correction parameter that captures direct dependency between the two scores. Subsequent research has also explored relaxing the independence assumption, with alternative approaches such as the bivariate Poisson model \citep{karlis2003analysis} and the Skellam-based goal-difference model \citep{karlis2009bayesian}. 
On the other hand, \citet{baio2010bayesian} adopt a Bayesian hierarchical framework and assume again \textit{conditional independence}, with the argument that dependence is introduced by the mixing of the variables at the higher level of the hierarchy. Following the arguments of \citet{Egidi2018}, who note that the correlation in seasonal leagues is small or absent \citep{McHale11}, we also relax the dependence assumption, allowing for a simpler formulation of the likelihood, and focusing on the interpretation and comparison of the CMP distribution with respect to the Poisson baseline.

In this context, our contribution to the existing literature lies in modelling departures from equidispersion in a robust and interpretable manner, while maintaining or improving predictive performance. Indeed, we show that football teams display systematic heterogeneity in their scoring behaviour that is not captured by the standard Poisson regression model's parameters. In the literature, the usage of the CMP distribution to model association football goals was first introduced by \citet{Piancastelli03042023}, as an example of a more general framework concerning multivariate CMP constructions through the Sarmanov method. A more direct application of the CMP distribution to a football model is recently studied by \citet{Florez24}, where the authors generalize the classical Poisson framework by adopting the CMP distribution with game-specific correlation random effect of the scores. Our proposed specification differs from this model in some key aspects. First, we adopt a more parsimonious parameterisation of the model, which enhances interpretability of the team-specific latent parameters and reduces the risk of over-parameterisation when each season is analysed independently. This modelling choice is motivated by the substantial variability between seasons in team composition and coaching staff, which may lead to notable changes in latent team characteristics, whereas \citet{Florez24} aggregate data across multiple seasons and primarily focus on inference related to the home advantage effect in the context of the COVID-19 period. Second, whereas \citep{Florez24} model dependence via correlation random effects, our applications to out-of-sample predictive tasks motivate a specification that avoids additional dependence structures that may not generalize well to unobserved matches. Finally, we explicitly model team-specific dispersion effects linked to goal-scoring behaviour through a SAS prior, and introduce a thresholding mechanism to distinguish between non-equidispersion and equidispersion, which can be particularly relevant in seasonal analyses, where the limited amount of observations can pose challenges to parameter identifiability. 

The proposed approach is implemented within a Bayesian framework with a tailored Metropolis-within-Gibbs sampler for efficient posterior exploration in the presence of SAS priors and correlated parameters. We apply the model on data from the last 5 seasons of the English Premier League (EPL), analysed independently, and demonstrate the improved model fit and out-of-sample predictive results of the CMP model with respect to the Poisson baseline. In each season, several teams are shown to depart from the equidispersed setting, with a propensity for over-dispersion in the EPL.

This paper is organized as follows. Section \ref{sec:CMP-SAS} details the CMP distribution, SAS priors and our newly developed model; Section \ref{sec:inference} describes details of the MCMC algorithm employed for parameter inference; Section \ref{ch:simulation_studies} showcases the results of the model on various simulated data settings; Section \ref{ch:applications} provides an example of the results on the 2023/24 season of the EPL, before extending the results to the last 5 years; Section \ref{ch:conclusions} contains a discussion on future work and concluding remarks.

\section{The Conway-Maxwell Poisson Spike-and-Slab Model}
\label{sec:CMP-SAS}
\subsection{The Conway-Maxwell Poisson Distribution} \label{section:compoisson_distribution}
The Conway-Maxwell-Poisson (CMP) is a discrete distribution on the non-negative integers, and can be seen as a two-parameter generalisation of the Poisson distribution \citep{ConwayMaxwell62}, allowing for greater flexibility by controlling for phenomena of \textit{non-equidispersion} in the observations. Using the parameterisation of \citet{guikema2008flexible}, the probability mass function for a COM-Poisson random variable \(Y\) with parameters \(\mu > 0\) and \(\nu \geq 0\) is defined over \(y \in \mathbb{N}^0\) as:
\begin{align}\label{eq:intractable_likelihood}
    f(y\mid\mu,\nu) &= \left(\frac{\mu^y}{y!}\right)^\nu \frac{1}{\sum_{x = 0}^{\infty} \left(\frac{\mu^x}{x!}\right)^\nu} = \frac{q_f(y\mid\mu,\nu)}{\mathcal{Z}_f(\mu,\nu)},
\end{align}
where the function is commonly decomposed in two parts: an unnormalized likelihood denoted \(q_f(y\mid\mu,\nu) = \left(\mu^y/y!\right)^\nu\), and an intractable normalizing constant denoted \(\mathcal{Z}_f(\mu,\nu) = \sum_{x = 0}^{\infty} \left(\mu^x/x!\right)^\nu\), independent of \(y\). The use of this distribution in statistical modelling has gained significant traction only in the recent decades, following the publication of the seminal article by \citep{Shmueli05}. This surge in interest can be largely attributed to the resolution of the challenges posed by the intractable normalizing constant, thanks to advancements in computing power and the development of statistical methods, particularly within the Bayesian framework.

The mode of \(f\) is \(\floor*{\mu}\), whereas its moments are generally unavailable in closed form, but their approximations can be obtained through the asymptotic representation of \(\mathcal{Z}_f(\mu,\nu)\) as \(\mathbb{E}(Y) \approx \mu + \frac{1}{2\nu} - \frac{1}{2}\) and \(Var(Y) \approx \frac{\mu}{\nu}.\)
The \(\nu\) parameter of the COM-Poisson controls for the \textit{dispersion} of the distribution. When \(\nu = 1\), we retrieve the (equidispersed) Poisson distribution as a special case:
\begin{align}\label{eq:pois_special_case}
    \mathcal{Z}_f(\mu, 1) = \sum_{y = 0}^{\infty} \frac{\mu^y}{y!} = e^\mu, \qquad    f(y\mid\mu, 1) = \frac{e^{-\mu}\mu^y}{y!}.
\end{align}
When \(\nu < 1\), the ratio will be smaller indicating heavier tails of the distribution, which is said to be \textit{overdispersed} (variance higher than expectation), while the opposite is true when \(\nu > 1\), resulting in an \textit{underdispersed} distribution (variance lower than expectation).
\subsection{Spike-and-Slab Priors}
Spike-and-slab (SAS) priors are a class of Bayesian mixture priors that combine two distinct components: a \textit{spike}, typically concentrated at a specific value, and a \textit{slab}, a diffuse distribution that allows parameters to vary over a wider range. This construction was originally developed in the context of variable selection of regression models, where the spike is centred at zero to represent exclusion of a predictor and the slab allows for non-zero effects of included predictors. 

The framework was initially developed by \citet{Mitchell1988BayesianVS}, and the method was further extended in seminal work by \citet{GeorgeMcCulloch1993} and \citet{IshwaranRao2005}. Formally, for a generic parameter \(\theta\), a typical SAS construction is:
\begin{align*}
    \theta \mid Z &\sim (1-Z)\ \underbrace{\delta_0(\theta)}_{\text{spike}} +\  Z \ \underbrace{f_{\text{diffuse}}(\theta)}_{\text{slab}},
\end{align*}
where \(\delta_0(\cdot)\) is the Dirac measure which concentrates the mass at \(0\), and \(Z\) is a latent binary indicator that selects between the spike (point mass at 0) and the slab (a diffuse distribution such as the Normal), typically modelled through a Bernoulli distribution. This type of specification with point mass mixture prior is typically defined as a discrete construction, as opposed to a continuous one which employs a concentrated continuous distribution over the spike prior \citep{TadesseVannucci2021book}.

While SAS priors are typically used for variable selection in regressions, the framework can be extended to an alternative perspective, where the spike represents a baseline or \textit{reference regime}, while the slab allows for deviations from that baseline in either direction. Similar perspectives have been discussed in the SAS literature, for instance by \citet{castillo2020spike} connecting it to multiple hypothesis testing, or by \citet{Rouder2018} as a Bayes factor approach. Posterior inference under such prior aims at simultaneously estimating both the magnitude of deviation from the baseline and the posterior mixing probabilities of the two regimes.

\subsection{Conway-Maxwell-Poisson Goal Model}
Our work extends the Poisson Goal model by generalizing the underlying likelihood with the CMP distribution. In particular, we propose \textit{team-specific} dispersion parameters \(\nu\), which affect the likelihood according to the \textit{attacking} team. In a match between teams \(i\) and \(j\), denote with \(y_{i,j}^{H}\) the score of team \(i\) (playing at home), and with \(y_{j,i}^{A}\) the score of team \(j\) (playing away). The likelihoods are modelled as follows:
\begin{align}
\begin{split} \label{eq:compoissonlikelihood_id}
    y_{i,j}^{H} \mid \mu_{i,j}^{H}, \nu_i &\sim \text{COM-Poisson}(\mu_{i,j}^{H}, \nu_{i}), \\
    y_{j,i}^{A} \mid \mu_{j,i}^{A}, \nu_j &\sim \text{COM-Poisson}(\mu_{j,i}^{A}, \nu_{j}), 
\end{split}
\end{align}
where the dispersion parameters \(\nu_i, \nu_j\) refer to the scores of team \(i\) and \(j\) respectively, and the centring parameters \(\mu_{i,j}^{H},\mu_{j,i}^{A}\) are as in the original log-linear parameterisation of \citet{Maher1982modelling}:
\begin{align}
\begin{split}\label{eq:log-linear_mu}
    log (\mu_{i,j}^{H}) &=   \alpha_{i} + \beta_{j} + \gamma, \\
    log (\mu_{j,i}^{A}) &=   \alpha_{j} + \beta_{i},
\end{split}
\end{align}
where \(\alpha_i, \beta_i\) refer respectively to the (log) attack and defence parameters of team \(i\). The distinction between home and away teams is made relevant by the \(\gamma\) parameter, representing the (log) home advantage, which is assumed identical across all teams in a given league consistently with the majority of the literature, e.g. \citet{Dixon1997}, \citet{baio2010bayesian}, and further motivated by the smaller sample sizes in seasonal league analysis. 

As previously mentioned in the introduction, some teams exhibit clear departures from equidispersion in their goal-scoring behaviour, whereas others appear to be well described by the equidispersed Poisson likelihood. Recall from Eq.[\ref{eq:intractable_likelihood}-\ref{eq:pois_special_case}] that whenever \(\nu = 1\), the CMP likelihood reduces to the (equidispersed) Poisson distribution as a special case. In light of this, we model the team-specific dispersion \(\nu_i\) using a SAS prior, introducing a binary indicator \(Z_i\) that determines the dispersion regime for team \(i\). Specifically, when \(Z_i = 0\), the spike component enforces \(\nu_i = 1\), corresponding to equidispersion, while when \(Z_i = 1\), the slab component assigns a diffuse prior mass around \(\nu_i = 1\), allowing for both over- and under-dispersion. Furthermore, we introduce an auxiliary parameter \(\eta_i \in \mathbb{R}\) that represents the log-state of the dispersion parameter \(\nu_i \in \mathbb{R}^+\), in order to enable unconstrained updates in the MCMC routine. 

Let \(N\) be the number of teams in the league. The priors on the team-specific dispersion parameters are specified hierarchically for \(i = 1,...,N\) as follows:
\begin{align}
    \pi(Z_i\mid p_i) &\sim \text{Bernoulli}(p_i), \nonumber\\
    \pi(p_i) &\sim \text{Beta}(a, b),\nonumber \qquad \ \ \ \ a, b >0,\\
    \pi(\eta_i) &\sim \text{Normal}(0, \sigma^2_\eta), \qquad \sigma_\eta > 0,\nonumber\\
    \nu_i &\coloneqq \exp(Z_i \ \eta_i), \label{eq:deterministic_nu_construction}
\end{align}
leading to the following behaviour for \(\nu_i\):
\begin{align*}\label{eq:nu_i_cases_sas}
    \nu_i = \begin{cases}
         1, &\text{if} \ Z_i = 0, \nonumber\\
        \exp(\eta_i), &\text{if} \ Z_i = 1. \nonumber
    \end{cases}
\end{align*}
This parameterisation induces a spike-and-slab mixture prior on \(\nu_i\):
\begin{align*}
    (\nu_i \mid Z_i) \sim (1-Z_i)\delta_1(\nu_i) + Z_i \ \text{LogNormal}(0, \sigma^2_\eta),
\end{align*}
where \(\delta_1(\cdot)\) is the Dirac measure concentrating all mass at \(1\), and the log-normal density is induced by \(\nu_i = \exp(\eta_i)\) when \(Z_i = 1\), since \(\eta_i\) is \textit{a priori} normally distributed.
To complete the model specification, we assign priors \(\pi(\cdot)\) over all the remaining parameters:
\begin{align*}
    \pi(\gamma) &\sim \text{Normal}(0, \sigma^2_\gamma), \qquad \sigma_\gamma > 0,\\
    \pi(\alpha_i) &\sim \text{Normal}(0, \sigma^2_\alpha), \qquad\sigma_\alpha > 0, \quad i = 1,\dots, N,\\
    \pi(\beta_i) &\sim \text{Normal}(0, \sigma^2_\beta), \qquad \sigma_\beta >0, \quad i = 1,\dots, N.
\end{align*}
Finally, hyperparameter values are set as
\begin{align*}
    a, b &= 1, \\
    \sigma_\gamma, \sigma_\alpha, \sigma_\beta &= 10, \\
    \sigma_\eta &= 1.
\end{align*}
The resulting priors on \(\gamma, \alpha_i\) and \(\beta_i\) are weakly informative, placing minimal constraints on their magnitudes. The slab prior for \(\eta_i\) is specified with a moderate scale to ensure most of the prior mass in plausible parameter values while maintaining clear separation from the spike at zero.

\section{Posterior Inference}
\label{sec:inference}
\begin{figure}[htbp]
  \centering
\begin{tikzpicture}
    \node [obs] (y_home) at (0,0) {\large $y_{i,j}^{\textcolor{myorange}{H}/\textcolor{myteal}{A}}$};
    \node [circle,draw=black,fill=white,inner sep=0pt,minimum size=0.8cm] (mu_home) at (-0.5,1.6) { $\mu_{i,j}^{H}$};

    \node [circle,draw=black,fill=white,inner sep=0pt,minimum size=0.8cm] (mu_away) at (0.5,1.6) { $\mu_{i,j}^{A}$};

    \node [circle,double, inner sep = 15pt, draw=black,fill=white,inner sep=0pt,minimum size=0.8cm] (nu_home) at (-3.5,0) { $\nu_{i}$};

    \node [circle, draw=black,fill=white,inner sep=0pt,minimum size=0.8cm] (phi_home) at (-5,1.6) { $\eta_{i}$};

    \node [circle, draw=black, fill=white, minimum size = 0.6cm] (Z_i) at (-3.5, 3.4) {$Z_i$};

    \node [circle, draw=black, fill=white, minimum size = 0.6cm] (p) at (-5, 3.4) {$p_i$};

    \node [circle, draw=black, fill=white, minimum size = 0.6cm] (att_i) at (-1.8, 3.4) {$\alpha_i$};

    \node [circle, draw=black, fill=white, minimum size = 0.6cm] (def_j) at (2.1, 3.4) {$\beta_j$};

    \node [circle, draw=black, fill=white, minimum size = 0.6cm] (home) at (-2.7, 5) {$\gamma$};

    \node [circle, minimum size=1cm, inner sep=0pt] (priors_theta) at (0, 5) {$ \sigma^2_\alpha,\sigma^2_\beta,\sigma^2_\gamma$};

    \node [text width=0.5cm] (sigma_nu) at (-6.4, 0.9) {$\sigma_{\eta}^2$};
    \node [circle, minimum size=0.5cm, inner sep=0pt] (p_prior) at (-5, 5) {$a,b$};

    \path [draw,->] (mu_home) edge[color = myorange] (y_home);
    \path [draw,->] (mu_away) edge[color = myteal] (y_home);
    \node[right] at ($(Z_i)!0.8!(nu_home)$) {\scriptsize{\(\nu_i = \exp(Z_i\eta_i)\)}};
    
    \path [draw,->] (nu_home) edge (y_home);
    \path [draw,->] (phi_home) edge[dashed] (nu_home);

    \path [draw,->] (sigma_nu) edge (phi_home);
    \path [draw,->] (sigma_nu) edge[dotted] (nu_home);

    \path [draw,->] (p) edge (Z_i);
    \path [draw,->] (Z_i) edge[dashed] (nu_home);

    \path [draw,->] (p_prior) edge (p);

    \path [draw,->] (att_i) edge (mu_home);
    \path [draw,->] (def_j) edge (mu_home);
    \path [draw,->] (home) |- (mu_home);
    \path [draw,->] (att_i) edge (mu_away);
    \path [draw,->] (def_j) edge (mu_away);

    \path [draw,->] (priors_theta) edge (att_i);
    \path [draw,->] (priors_theta) edge (def_j);
    \path [draw,->] (priors_theta) edge (home);
    \plate [color=blue] {part1} {(y_home)(mu_home)(att_i)(mu_away)(nu_home)(phi_home)(Z_i)} {};
    \plate [color=red, yshift = 0.1cm] {part2} {(y_home)(mu_home)(mu_away)(def_j)(part1.north east)} {};
    \node[anchor=south west, color = blue] at (part1.south west) {$i = 1,...,N$};
    \node[anchor=south east, color = red] at (part2.south east) {$j \neq i$};
\end{tikzpicture}
\caption{Graphical model illustrating the parameter dependency structure of the observed data \(\bm{Y}\).}
  \label{fig:graphical-model}
\end{figure}
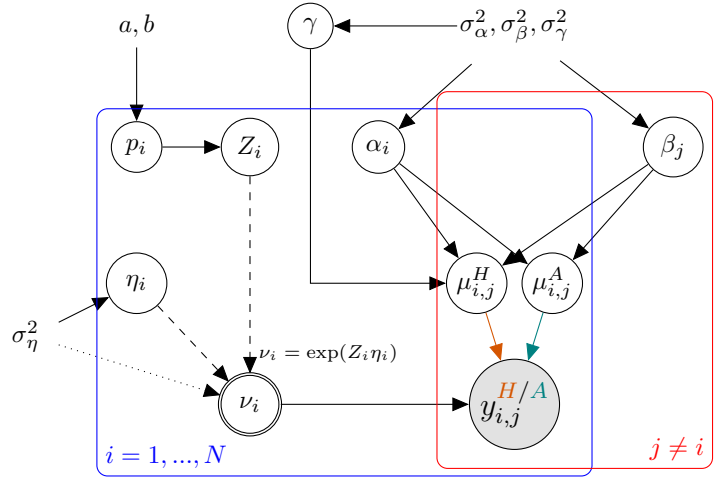

The joint target posterior distribution of interest is:
\begin{align*}
    P&(\bm{\alpha},\bm{\beta}, \gamma, \bm{\eta}, \bm{Z}, \bm{p}\mid \bm{Y}) \propto \nonumber \\
    &\prod_{i=1}^{N} \prod_{j\neq i} \left[f_{CMP}(y_{i,j}^H\mid \mu_{i,j}^H, \nu_i)f_{CMP}(y_{i,j}^A\mid \mu_{i,j}^A, \nu_i) \right]\pi(\bm{\alpha}) \pi(\bm{\beta}) \pi(\gamma) \pi(\bm{\eta}) \pi(\bm{Z}\mid \bm{p}) \pi(\bm{p}),
\end{align*}
where the posterior factorisation (up to proportionality) is implied by the conditional independencies among parameters, as encoded in the graphical model of Figure~\ref{fig:graphical-model}. We exploit this factorisation to perform Gibbs sampling, iteratively sampling each parameter from its full conditional distribution. For the parameters that appear directly in the likelihood function, the full conditional is not available in closed form, since the normalizing constant renders the likelihood intractable (as noted in Section \ref{section:compoisson_distribution}). In the following section, we describe a sampling strategy that circumvents the need to evaluate this constant.
\subsection{Doubly-intractable Likelihood}\label{sec:intractable_likelihood}
Denote \(\bm{\theta} = (\mu,\nu)\) the set of parameters of the COM-Poisson for notational simplicity. As is usual in the Bayesian framework, we are interested in studying the posterior distribution of the parameters \(\bm{\theta}\), given a sequence of observations \(\bm{y} = y_1,...,y_n\):
\begin{align*}
    p(\bm{\theta}|\bm{y}) = \frac{f(\bm{y}|\bm{\theta})p(\bm{\theta})}{\int f(\bm{y}|\bm{\theta})p(\bm{\theta})d\bm{\theta}} = \frac{q_f(\bm{y}|\bm{\theta})p(\bm{\theta})}{\mathcal{Z}_f(\bm{\theta})\int f(\bm{y}|\bm{\theta})p(\bm{\theta})d\bm{\theta}}.
\end{align*}

The posterior \(p(\bm{\theta}|\bm{y})\) is said to be \textit{doubly-intractable} because of the two intractable terms: the first is the usual posterior model evidence \(  \int f(\bm{y}|\bm{\theta})p(\bm{\theta})d\bm{\theta} \), and the second is the intractable normalising constant of the CMP likelihood \(\mathcal{Z}_f\). Simulation methods such as Markov Chain Monte Carlo (MCMC) are one of the ways to circumvent the first intractability, by generating samples of the parameters which asymptotically converge to the posterior distribution. The classic Metropolis-Hastings (MH) algorithm generates a proposal \(\bm{\theta}^*\) from the current set of parameters \(\bm{\theta}\) through a proposal distribution \(k(\bm{\theta}^*\mid\bm{\theta})\), and accepts the proposed values as the next set of parameters in the chain with probability \(a(\bm{\theta}, \bm{\theta}^*)\):
\begin{align*}
    a(\bm{\theta}, \bm{\theta}^*) &= \min{\left\{ 1, \frac{f(\bm{y}\mid\bm{\theta}^*)}{f(\bm{y}\mid\bm{\theta})}\frac{k(\bm{\theta}\mid\bm{\theta}^*)}{k(\bm{\theta}^*\mid\bm{\theta})}\frac{\pi(\bm{\theta}^*)}{\pi(\bm{\theta})}\right\}} =\min{\left\{1, \frac{\frac{q_f(\bm{y}\mid\bm{\theta}^*)}{\mathcal{Z}_f(\bm{\theta}^*)}}{\frac{q_f(\bm{y}\mid\bm{\theta})}{\mathcal{Z}_f(\bm{\theta})}}\frac{k(\bm{\theta}\mid\bm{\theta}^*)}{k(\bm{\theta}^*\mid\bm{\theta})}\frac{\pi(\bm{\theta}^*)}{\pi(\bm{\theta})}\right\}},
\end{align*}
which, for our case, is intractable due the presence of the (sequence) of normalizing constant(s) \(\mathcal{Z}_f\). To tackle this problem, we follow the approach of \citet{Benson21} to sample parameter values for the CMP distribution, which is centred around the \textit{Exchange} algorithm \citep{Moller06, Murray12}. This method augments the posterior distribution with auxiliary variables simulated from the likelihood under the proposed parameters, \(\bm{y}' \sim f(\cdot \mid \bm{\theta}^*)\). The acceptance ratio in the augmented posterior becomes: 
\begin{align*}
    a(\bm{\theta}, \bm{\theta}^*) 
    &= \min{\left\{
    1, \frac{\frac{q_f(\bm{y}|\bm{\theta}^*)}{\mathcal{Z}_f(\bm{\theta}^*)}}{\frac{q_f(\bm{y}|\bm{\theta})}{\mathcal{Z}_f(\bm{\theta})}}\frac{\frac{q_f(\bm{y}'|\bm{\theta})}{\mathcal{Z}_f(\bm{\theta})} k(\bm{\theta}|\bm{\theta}^*)}{\frac{q_f(\bm{y}'|\bm{\theta}^*)}{\mathcal{Z}_f(\bm{\theta}^*)} k(\bm{\theta}^*|\bm{\theta})}\frac{p(\bm{\theta}^*)}{p(\bm{\theta})}
    \right\}} \nonumber= \min{\left\{
    1, \frac{q_f(\bm{y}|\bm{\theta}^*)}{q_f(\bm{y}|\bm{\theta})}\frac{q_f(\bm{y}'|\bm{\theta})} {q_f(\bm{y}'|\bm{\theta}^*)}\frac{k(\bm{\theta}|\bm{\theta}^*)}{k(\bm{\theta}^*|\bm{\theta})}\frac{p(\bm{\theta}^*)}{p(\bm{\theta})}\right\}},
\end{align*}
where the evaluation of the auxiliary draws causes the normalizing constants \(\mathcal{Z}_f\) to cancel out, thus allowing full tractability of the ratio. Further details are provided in \ref{app:exchange_mcmc}, and the exact draws from the likelihood can be efficiently obtained through a Rejection sampler in \ref{app:A_rej_sampler}.

Furthermore, \citet{Benson21} also describe a procedure for constructing \textit{an unbiased estimator} for the CMP likelihood, which is useful for likelihood-based model evaluation tasks, such as the computation of information criteria. They propose a method to create an estimate of the intractable likelihood by leveraging the number of draws \(N_r\) that are required to obtain \(r\) acceptances in the rejection sampler. In fact, this value may be seen as a measure of how closely the envelope matches the target distribution. More specifics on the estimator can be found in Appendix \ref{appendix:llhood_estimator}.

\subsection{Identifiability of the Attack and Defence Parameters}
In order to implement the models, we need to estimate the team-specific parameters for each of the \(N\) teams. This is done in the literature through MLE or Bayesian approaches, by finding the optimal solution for the set of equations given by the scoring rate parameters in Eq.[\ref{eq:log-linear_mu}], for all given pairs \(i,j\). Crucially, this set of equations is not identifiable since we can add and subtract the same constant to each attack and defence parameter, resulting in infinitely many equivalent solutions of the optimisation problem. In order to identify a unique solution to the problem, following e.g. \citet{baio2010bayesian}, we impose a \textit{sum-to-zero} constraint, where we fix the respective sums of parameters for all teams to be equal to 0:
\begin{equation}\label{eq:zerosum_constraint}
    \sum_{i=1}^{N} \alpha_i = 0, \qquad  \sum_{i=1}^{N} \beta_i = 0.
\end{equation}
To improve the mixing and exploration of the parameter space in our MCMC routine, we update each team's parameters \textit{individually}. In order to enforce the constraint in Eq.[\ref{eq:zerosum_constraint}] while preserving symmetry in the proposal distribution, we work within a \(N-1\) hyperplane with the unconstrained updates:
\begin{equation}\begin{aligned}\label{eq:alpha_beta_rwprop}
     \alpha^*_i \sim k(\alpha_i^* \mid \alpha_i) &= \text{Normal}(\alpha_i,s^2_{\alpha}), \\
     \beta_i^* \sim k( \beta_i^* \mid \beta_i ) &= \text{Normal}(\beta_i, s^2_{\beta}),
\end{aligned}
\hspace{1cm} s_{\alpha}, s_{\beta} >0\ , \quad i = 1, \dots, N-1,
\end{equation}
where we use the notation \(s\) to denote the standard deviation of the proposal distributions, and satisfy the constraint by imposing, at each proposal of \(\alpha_i^*\) and \(\beta_i^*\), the sum-to-zero constraint on a pre-determined parameter:
\begin{align}\label{eq:stz_constraint}
     \alpha^*_N \coloneqq -\left( \alpha_i^* + \sum_{\substack{j=1 \\ j \neq i}}^{N-1} \alpha_j\right), \hspace{3cm} \beta^*_N \coloneqq -\left( \beta_i^* 
     + \sum_{\substack{j=1 \\ j \neq i}}^{N-1} \beta_j\right),
\end{align}
where, without loss of generality and for simplicity of exposition, we set the \textit{constrained} team to be the last one with index \(N\). Note that changing which team is constrained did not result in meaningful changes of posterior estimates in our experiments.

\subsection{Correlated \(\alpha_i\) and \(\eta_i\) Updates}\label{sec:correlation_att_nu}
\begin{figure}[tbp]
    \centering
    \includegraphics[width=0.9\linewidth]{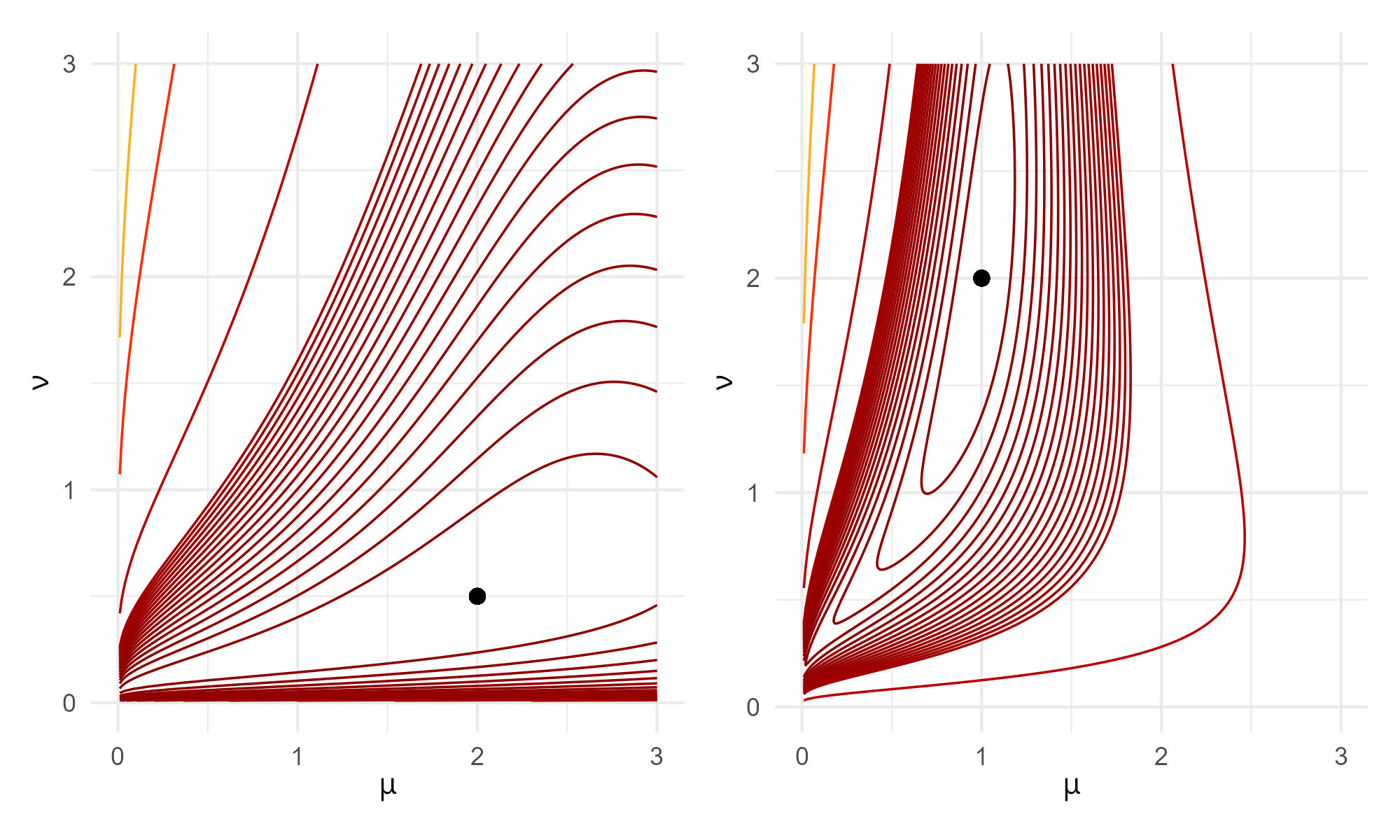}
    \caption{Log-likelihood contour of a CMP distribution for varying values of its parameters, based on observations generated from a CMP\((\mu = 2, \nu = 0.5)\) (over-dispersed, left plot) and \((\mu = 1, \nu = 2)\) (under-dispersed, right plot). Darker levels indicate higher log-likelihood, and black dot indicates true parameters and max log-likelihood.}
    \label{fig:contour}
\end{figure}

Within the region of the parameter space that is reasonable for our application, the optimal CMP parameters \((\mu, \nu)\) are positively correlated, as evidenced by the geometry of the log-likelihood contours shown in Figure~\ref{fig:contour}. 
In our modelling context, this translates to a correlation between \(\alpha_i\) and \(\eta_i\) for each team \(i\). To capture this dependence efficiently in the MCMC, we couple the proposal of these two parameters through a bivariate Gaussian distribution with positive correlation coefficient \(\rho\), common for all teams \(i = 1,...,N\), and accept or reject these two parameters jointly. In practice, the proposal for \(i = 1,...,N-1\) are as follows:
\begin{align}\label{eq:biv_proposal}
    k(\alpha_i^*,\eta_i^* \mid \alpha_i, \eta_i) = \text{MV-Normal}\left(\begin{bmatrix}
\alpha_i\\
\eta_i
\end{bmatrix}, \bm{\Sigma} = \begin{bmatrix}
s^2_\alpha & \rho \ s_\alpha s_\eta\\
\rho \ s_\alpha s_\eta & s^2_\eta
\end{bmatrix}\right).
\end{align}
For the constrained parameter \(\alpha_N\), the proposal parameter \(\alpha_N^*\) is deterministic given the other updates (Eq.[\ref{eq:stz_constraint}]). We can generate an \textit{induced} conditional proposal for \(\eta_N^*\) from the same joint proposal density as in Eq.[\ref{eq:biv_proposal}] for the pair \(\{\alpha_N^*, \eta_N^*\}\) through the conditional distribution of a bivariate Normal distribution:
\begin{align}\label{eq:fixnu_proposal}
    \Delta\alpha_N &= \alpha_N^* - \alpha_N, \nonumber \\ 
    k(\eta_N^* \mid \alpha_N^*) &\sim \text{Normal}\left(\eta_N + \rho\frac{s_\eta}{s_\alpha}\Delta\alpha_N, (1-\rho^2)s^2_\eta\right).
\end{align}
Lastly, proposal standard deviations and correlation coefficient for our application are selected as follows:
\begin{align*}
    s_\alpha, s_\beta = 0.10, \qquad s_\gamma &= 0.08, \qquad s_\eta = 0.4,\\
    \rho &= 0.85,
\end{align*}
where these values were chosen to obtain optimal acceptance rates according to the MH literature, e.g., \citep{Roberts1997WeakCA,Neal_2006}.

\subsection{Metropolis-within-Gibbs Algorithm}\label{sec:mwgb_algorithm}
We now combine all the elements described in the previous sections into a Metropolis-within-Gibbs sampler (MWGS). The closed form full conditional for \(\bm{p}\) is sampled directly, while parameters present directly in the likelihood are handled via MH updates using the exchange algorithm. The parameter \(\bm{\nu}\) is not included in the MWGS algorithm since it is deterministically retrieved through the \(\bm{Z}\) and \(\bm{\eta}\) parameters (Eq.[\ref{eq:deterministic_nu_construction}]); at each iteration \textit{t} of the algorithm, the state space is represented by \(\bm{\theta}^{(t)} = \{\bm{\alpha}^{(t)}, \bm{\beta}^{(t)}, \gamma^{(t)},\bm{\eta}^{(t)},\bm{Z}^{(t)}, \bm{p}^{(t)}\}\). 

In the MCMC routine, we update the parameters for each team \(i\) individually: this improves mixing and allows for more efficient posterior exploration. To facilitate notation in the algorithm description, we introduce compact representations of some relevant quantities as follows. The notation \(\bm{\phi}_{-i}\) indicates the vector \(\{\bm{\phi} \setminus \phi_i\}\), i.e. the full set of parameters excluding the one indexed by \(i\). We write \(\bm{Y}_{i,\cdot} = \{y_{i,j}^H,y_{i,j}^A\}_{j \neq i}\) to denote the set of \textit{all} the goals \textit{scored} by team \(i\) against all the other teams, both at home or away. Conversely, we use the notation \(\bm{Y}_{\cdot, i} = \{y_{j,i}^H, y_{j,i}^A\}_{j \neq i}\) to indicate the set of \textit{all} goals \textit{conceded} by team \(i\). Furthermore, iteration indices \(t\) are suppressed. An asterisk denotes a proposed value, \(\phi_i^*\), while the absence of an asterisk denotes the current (i.e., most recently accepted) value, \(\phi_i\). 

In the following, we include the main computational steps of the MCMC algorithm, for a full detailed report with full conditionals refer to Appendix \ref{app:full_conditionals}. %
\begin{itemize}
    \item For the latent assignments \(Z_i\), the full conditional is not available in closed form given the intractability of the likelihood discussed in Section \ref{sec:intractable_likelihood}. We apply a MH step with Exchange and propose \(Z_i^*\) by binary flip of the current state, i.e. \(Z_i^* = 1 - Z_i\). Denote the current state space with \(\bm{\theta}\), and the state space with the proposal \(Z^*\) as \(\bm{\theta}^*\). Draw auxiliary data \(\bm{Y_{i, \cdot}'} = \{y_{i,j}^{H'}, y_{i,j}^{A'}\}_{j \neq i} \sim \text{CMP}(\bm{\theta}^*)\), and compute the acceptance ratio from \(Z_i\) to \(Z_i^*\):
        \begin{align}\label{eq:Z_acceptance_ratio}
            A_Z = \min \Biggl\{1, \frac{
            \prod_{j \neq i} q(y_{i,j}^H, y_{i,j}^A \mid \bm{\theta^*})q(y_{i,j}^{H'}, y_{i,j}^{A'} \mid \bm{\theta})
            }{
            \prod_{j \neq i} q(y_{i,j}^H, y_{i,j}^A \mid \bm{\theta} ) q(y_{i,j}^{H'}, y_{i,j}^{A'} \mid \bm{\theta}^*)
            } \frac{f_{\text{Bern}}(Z_i^* \mid p_i)}{f_{\text{Bern}}(Z_i \mid p_i)}
            \Biggr\},
        \end{align}
        where the proposal densities \(k(\cdot \mid \cdot) \), the normalizing constants \(\mathcal{Z(\bm{\theta})}, \mathcal{Z}(\bm{\theta}^*)\) and priors on \(p_i\) cancel out and are directly omitted. With probability \(A_Z\), accept the proposal \(Z^*_i\) and set \(Z_i \coloneqq Z_i^*\). Note that if proposing \(Z_i^* = 0\), the auxiliary draws corresponds to samples from a Poisson distribution.\\
        
\item For the hierarchical assignment probabilities \(\bm{p}\), we can directly perform a Gibbs update by drawing \(p_i\) independently across \(i\):
\begin{align}
\bm p \mid \bm Z \;\sim\; \prod_{i=1}^N \text{Beta}\!\left(\alpha_p + Z_i,\; \beta_p + 1 - Z_i\right).
\label{eq:p_block_draw_compact}
\end{align}
\item The latent dispersion \(\eta_i\) and attack \(\alpha_i\) parameters are positively correlated and as such are updated jointly, as explained in Section \ref{sec:correlation_att_nu}. Additionally, the change in \(\alpha_i\) induces a deterministic change in \(\alpha_N\) from the constraint in Eq.[\ref{eq:stz_constraint}], influencing the terms \(\bm{Y}_{N \cdot}\) of the likelihood. To reflect this change in \(\alpha_N\), \(\eta_N\) is jointly proposed. We employ a MH step with positively correlated bivariate proposal distribution centred at the previous state according to Eq. [\ref{eq:biv_proposal}] to generate \(\{\alpha_i^*, \eta_i^*\}\), and the corresponding induced proposal from Eq.[\ref{eq:fixnu_proposal}] to retrieve \(\{\alpha_N^*, \eta_N^*\}\). Denote the current state space with \(\bm{\theta}\), and the state space with the proposal \(\{\alpha_i^*, \alpha_N^*, \eta_i^*, \eta_N^*\}\) as \(\bm{\theta}^*\). Draw auxiliary data \(\bm{Y_{i, \cdot}'}, \bm{Y}_{N ,\cdot}' \sim \text{CMP}(\bm{\theta}^*)\). The resulting parameters are accepted with probability \(a(\{\alpha_i, \alpha_N, \eta_i, \eta_N\}, \{\alpha_i^*, \alpha_N^*, \eta_i^*, \eta_N^*\})\) equal to:
\begin{align}
\footnotesize
\begin{split}\label{eq:mh_acc_alpha_eta}
A_{\alpha,\eta} = \min \Biggl\{1, &\frac{\prod_{j\neq i}q(y_{i,j}^{H}, y_{i,j}^{A} \mid \bm{\theta}^*) \, q(y_{i,j}^{H'}, y_{i,j}^{A'} \mid \bm{\theta})}{
\prod_{j\neq i} q(y_{i,j}^{H}, y_{i,j}^{A} \mid \bm{\theta})\,q(y_{i,j}^{H'}, y_{i,j}^{A'} \mid \bm{\theta}^*)
}\frac{f_{\mathcal{N}}(\alpha_i^* \mid 0, \sigma_\alpha^2)\,f_{\mathcal{N}}(\eta_i^* \mid 0, \sigma_\eta^2)}{f_{\mathcal{N}}(\alpha_i\mid 0, \sigma_\beta^2)\,f_{\mathcal{N}}(\eta_i \mid 0, \sigma_\eta^2)}\\
& \qquad \qquad \times \frac{\prod_{j\neq N} q(y_{N,j}^{H}, y_{N,j}^{A} \mid \bm{\theta}^*)\,q(y_{N,j}^{H'}, y_{N,j}^{A'} \mid \bm{\theta})}{
\prod_{j\neq N} q(y_{N,j}^{H}, y_{N,j}^{A} \mid \bm{\theta})\,q(y_{N,j}^{H'}, y_{N,j}^{A'} \mid \bm{\theta}^*)
}\frac{f_{\mathcal{N}}(\eta_N^* \mid 0, \sigma_\eta^2)}{f_{\mathcal{N}}(\eta_N \mid 0, \sigma_\eta^2)}\Biggr\},
\end{split}
\normalsize
\end{align}%
where we omit the symmetric proposal kernels \(k(\cdot \mid \cdot) \) and the normalizing constants \(\mathcal{Z(\bm{\theta})}, \mathcal{Z}(\bm{\theta}^*)\) as they cancel out. The prior evaluations of \(\alpha_N, \alpha_N^*\) are not included given their change is deterministic.\\
\item For the defence parameters \(\beta_i\), the observations of interest will be the goals \textit{conceded} by team \(i\), \( \bm{Y}_{\cdot, i}\). Similarly to the attack case, changes to \(\beta_i\) induce changes to \(\beta_N\). We generate a proposal defence parameter \(\beta_i^*\) and the induced change in \(\beta_N^*\) according to Eqs.[\ref{eq:alpha_beta_rwprop}-\ref{eq:stz_constraint}]. Define the current state space with \(\bm{\theta}\), and with \(\bm{\theta}^*\) the state space with the proposal parameters \(\{\beta_i^*, \beta_N^*\}\), and sample auxiliary data \(\bm{Y}_{\cdot, i}', \bm{Y}_{\cdot, N}' \sim \text{CMP}(\bm{\theta}^*)\). The resulting acceptance probability of transitioning from \(\{\beta_i, \beta_N\}\) to \(\{\beta_i^*, \beta_N^*\}\) is:
\begin{align}\label{eq:mh_acc_def}
\footnotesize
\begin{split}
A_\beta = \min \Biggl\{1&,\frac{ 
\prod_{j\neq i} q(y_{j,i}^{H}, y_{j,i}^{A} \mid \bm{\theta}^*)\,q(y_{j,i}^{H'}, y_{j,i}^{A'} \mid \bm{\theta})}{
\prod_{j\neq i} q(y_{j,i}^{H}, y_{j,i}^{A} \mid \bm{\theta})\,q(y_{j,i}^{H'}, y_{j,i}^{A'} \mid \bm{\theta}^*)
}\frac{f_{\mathcal{N}}(\beta_i^* \mid 0, \sigma_\beta^2)}{f_{\mathcal{N}}(\beta_i\mid 0, \sigma_\beta^2)} \\
&\qquad \qquad \qquad\times \frac{
\prod_{j\neq N} q(y_{j,N}^{H}, y_{j,N}^{A} \mid \bm{\theta}^*)\,q(y_{j,N}^{H'}, y_{j,N}^{A'} \mid \bm{\theta})
}{
\prod_{j\neq N} q(y_{j,N}^{H}, y_{j,N}^{A} \mid \bm{\theta})\,q(y_{j,N}^{H'}, y_{j,N}^{A'} \mid \bm{\theta}^*)
} \Biggr\}
\end{split}
\normalsize
\end{align}%
where we omit the symmetric proposal kernels \(k(\cdot \mid \cdot) \), and the normalizing constants \(\mathcal{Z(\bm{\theta})}, \mathcal{Z}(\bm{\theta}^*)\) as they cancel out.\\
\item Lastly, for the home coefficient \(\gamma\), we only need to consider the goals scored by the home teams, \(\bm{Y}^H = \{y_{i,j}^H\}_{i = 1, j \neq i}^N\), and an analogous intractability arises. We proceed in a similar fashion, with a Gaussian proposal:
\begin{align}\label{eq:gamma_proposal}
    \gamma^* \sim k(\gamma^* \mid \gamma) =  \text{Normal}(\gamma, s^2_{\gamma}).
\end{align}
We denote the current state space with \(\bm{\theta}\), and the proposal state space by \(\bm{\theta}^*\). Sample \(\bm{Y}^{H'} \sim \text{CMP}(\bm{\theta}^*)\), and evaluate the transition probability from \(\gamma\) to \(\gamma^*\):
\begin{align}\label{eq:mh_acc_home}
    A_\gamma = \min\left\{1, \frac{
    \prod_{i = 1}^N \prod_{j\neq i} q(y_{i,j}^H \mid \bm{\theta}^*) q(y_{i,j}^{H'} \mid \bm{\theta})}{
    \prod_{i = 1}^N \prod_{j\neq i} q(y_{i,j}^H \mid \bm{\theta}) q(y_{i,j}^{H'} \mid \bm{\theta}^*)
    } \frac{f_\mathcal{N}(\gamma^* \mid 0, \sigma^2_\gamma)
    }{f_\mathcal{N}(\gamma \mid 0, \sigma^2_\gamma)
    }\right\}.
\end{align}
\end{itemize}
Algorithm [\ref{alg:MWGS}] succinctly summarizes the computational steps highlighted in this section to obtain samples from the CMP-SAS model. Note that we can retrieve a slab-only CMP model (CMP-Full) by enforcing \(Z_i = 1\) for all \(i\), and the Poisson model by setting \(Z_i = 0\) for all \(i\). 
\begin{algorithm}[h]\label{alg:MWGS_CMP-SAS}
\begin{minipage}{\linewidth} 
\setstretch{1.25}
\caption{MWGS for the CMP-SAS Model}\label{alg:MWGS}
\KwIn{Data \(\bm{Y}\), n. of teams \(N\), iterations \(T\), starting values \(\{\bm{\alpha}^{(1)}, \bm{\beta}^{(1)}, \gamma^{(1)},\bm{\eta}^{(1)},\bm{Z}^{(1)}, \bm{p}^{(1)}\}\).}
\For{\(t \ \textbf{in} \ 2,…,T\)}{
    \For{\(i \ \textbf{in} \ 1,…,N\)} {%
        Propose \(Z_i^* = 1- Z_i\), \\
        sample auxiliary data \(\bm{Y}_{i, \cdot}' \sim \text{CMP}(\bm{\theta}^*)\), \\
        set \(Z_i \coloneqq Z_i^*\) with probability \(A_Z\) computed from Eq.[\ref{eq:Z_acceptance_ratio}]\;\vspace{0.1cm}
        Sample \(p_i\) using the conjugate full conditional Eq.[\ref{eq:p_block_draw_compact}].
        }
        \For{\(i\) \textbf{in} \(1,\dots, N-1\)}{
            Generate proposals \(\alpha_i^*, \eta_i^*\) with Eq.[\ref{eq:biv_proposal}], compute \(\alpha_N^*\) and sample \(\eta_N^*\) using Eqs.[\ref{eq:stz_constraint}-\ref{eq:fixnu_proposal}], \\
            sample auxiliary data \(\bm{Y}_{i,\cdot}', \bm{Y}_{N,\cdot}'\sim CMP(\bm{\theta}^*)\), \\
            set \(\{\alpha_i, \alpha_N, \eta_i, \eta_N\} \coloneqq \{\alpha_i^*, \alpha_N^*, \eta_i^*, \eta_N^*\}\) with probability \(A_{\alpha,\eta}\) from Eq.[\ref{eq:mh_acc_alpha_eta}]\; \vspace{0.1cm}
            Generate proposals \(\beta^*_i, \beta_N^*\) through Eqs.[\ref{eq:alpha_beta_rwprop}-\ref{eq:stz_constraint}], \\
            sample auxiliary data \(\bm{Y}_{\cdot, i}',\bm{Y}_{\cdot, N}'\sim CMP(\bm{\theta}^*)\),\\
            set \(\{\beta_i, \beta_N\} \coloneqq \{\beta_i^*, \beta_N^*\}\) with probability \(A_\beta\) from Eq.[\ref{eq:mh_acc_def}]\;
        }
        Generate proposal \(\gamma^*\) with Eq.[\ref{eq:gamma_proposal}], \\
        sample auxiliary data \(\bm{Y}^{H'}\sim CMP(\bm{\theta}^*)\),\\ 
        set \(\gamma \coloneqq \gamma^*\) with probability \(A_\gamma\) from Eq.[\ref{eq:mh_acc_home}]\;\vspace{0.1cm}

        Set \(\{\bm{\alpha}^{(t)}, \bm{\beta}^{(t)}, \gamma^{(t)},\bm{\eta}^{(t)},\bm{Z}^{(t)}, \bm{p}^{(t)}\} \coloneqq \{\bm{\alpha}, \bm{\beta}, \gamma, \bm{\eta}, \bm{Z}, \bm{p}\}.\)
    }
\KwRet{\(\{\bm{\alpha}^{(t)}, \bm{\beta}^{(t)}, \gamma^{(t)},\bm{\eta}^{(t)},\bm{Z}^{(t)}, \bm{p}^{(t)}\}_{t = 1}^T\)}
\end{minipage}
\end{algorithm}%
\vspace{-0.8cm}
\subsection{Model Evaluation}
\subsubsection{WAIC}
Many measurement methods are available in the literature to compare model fit. To estimate the predictive accuracy of our in-sample experiments, we will be employing the \textit{Widely Applicable Information Criterion} (WAIC) \citep{watanabe2010waic}, as it is a fully Bayesian method that leverages the whole posterior distribution, as opposed to other metrics such as the \textit{Deviance Information Criterion} (DIC) \citep{Spiegelhalter02DIC}, which is less preferable when the posterior distribution is not well summarized by its mean \citep{gelman2013waic}. This is precisely our case, given the multi-modal nature of the dispersion parameters \(\nu_i\) arising from the spike-and-slab construction. The WAIC is computed from the log point-wise predictive density (lppd) and a penalisation term that accounts for the model complexity:
\begin{align}\label{eq:waic}
    \text{lppd} &= \sum_{k = 1}^n \log \left(\frac{1}{S}\sum_{s = 1}^{S} f(y_k \mid \bm{\theta}^{(s)})\right)\nonumber,\\
    p_{\text{WAIC}} &=  \sum_{k = 1}^n \text{Var}_\theta\big[\log \big(f(y_k \mid \bm{\theta}
    )\big)\big]\nonumber,\\[1ex]
    \text{WAIC} &= -2(\text{lppd} - p_{\text{WAIC}}),
\end{align}
where the predictive densities are computed for \(S\) samples from the posterior, and averaged over the number of observations \(n\). In order to compute the point-wise log-likelihood, we can leverage an unbiased estimator of the likelihood, previously mentioned in Section \ref{sec:intractable_likelihood} and detailed in Appendix \ref{appendix:llhood_estimator}.

\subsubsection{Posterior Predictive}\label{ch:posterior_predictive}
Aside from information criteria, which \textit{estimate} the expected predictive performances of the models, the \textit{out-of-sample} observations are directly computable and are arguably of larger interest in the context of football. As usual in the Bayesian framework, prediction of new observations \(\tilde{y}\) may be performed directly through the posterior predictive distribution:
    \begin{equation*}
        p(\tilde{y}\mid y) = \int_{\Theta} p(\tilde{y},\bm{\theta}\mid y)d\bm{\theta} = \int_{\Theta} p(\bm{\theta}\mid y)p(\tilde{y}\mid\bm{\theta})d\bm{\theta},
    \end{equation*}
where the integral is estimated by averaging over \(S\) posterior samples. To evaluate the out-of-sample predictive distributions, we estimate the parameters using only the data up to a certain point of the season. 

In particular, starting from half of the season, we predict the following week worth of games, which in a typical football league of \(20\) teams corresponds to \(10\) matches, where each team plays one game (usually in the same day or weekend). We then infer the parameters with the addition of these new games and repeat this cycle until the end of the season. In a league of \(N = 20\) teams, we start predicting after \(190\) games, resulting in a total of \(19\) training sessions.

In order to evaluate and compare the different models' predictive performance, several metrics are potentially available, and the literature is not in agreement in which is the most appropriate. In their paper, \citet{constantinou2012solving} argue that the \textit{Rank Probability Score} (RPS), introduced by \citet{Epstein69}, is the most appropriate for the \textit{three-way probabilistic forecasts} of football outcomes. More recently, \citet{wheatcroft2021evaluating} argues against the examples used by \citet{constantinou2012solving}, claiming that the settings used were overly simplistic. Instead, the author advocates for the use of a \textit{local} scoring rule instead, the \textit{ignorance} score:
\begin{align}\label{eq:ignorance_score}
    \text{IGN} = -\log_2(\tilde{f}(y^*)),
\end{align}
 where \(y^*\) represents the actual observed outcome. According to the author, both metrics are \textit{proper} in retrieving the correct optimal predictive model, although the Ignorance score is claimed to be more \textit{efficient}. Without entering into the details of this discourse, we use the IGN score for predictive comparison, and note that results using the RPS lead to similar conclusions on the predictive performance comparisons.
 
\section{Simulation Studies}\label{ch:simulation_studies}
To investigate the performance of the proposed CMP model with spike-and-slab prior  (\textit{CMP-SAS}), we conduct a set of simulation studies to assess both its parameter recovery capabilities and its comparative performance with respect to the baseline Poisson model and an alternative CMP model without the spike-and-slab indicators (\textit{CMP-Full}). In the first section, we investigate how well the CMP model captures the underlying dispersion structure by generating data under different levels of non-equidispersion. In the second section, we assess the overall model fit by comparing the CMP models against the baseline Poisson model using the Widely Applicable Information Criterion (WAIC).
\subsection{Dispersion Recovery}\label{sec:sim_dispersion_recovery}
In this subsection, we examine how well the CMP-SAS model can recover the underlying dispersion parameters when data are generated from the model itself. The settings we use are designed to challenge the model's capacity to identify departures from equidispersion and to assess the accuracy and consistency of parameter estimates. In particular, we analyse the scenarios where the true dispersions are \textit{over-dispersed} \(\nu_i^{true} < 1\) and \textit{under-dispersed} \(\nu_i^{true} > 1\). To this end, we generate synthetic leagues that mimic realistic parameters of football leagues with \(N = 20\) teams, each containing \(10\) equidispersed teams (\(\nu_i^{true} = 1\)), and \(10\) non-equidispersed teams \((\nu_i^{true} =k\neq 1)\), with \(0.2\leq k\leq 4\), reflecting a wide range of dispersions. For each value of the true dispersion \(k\), we replicate \(5\) synthetic leagues with different random seeds.

\begin{figure}[b!p]
\centering\includegraphics[width=0.95\linewidth]{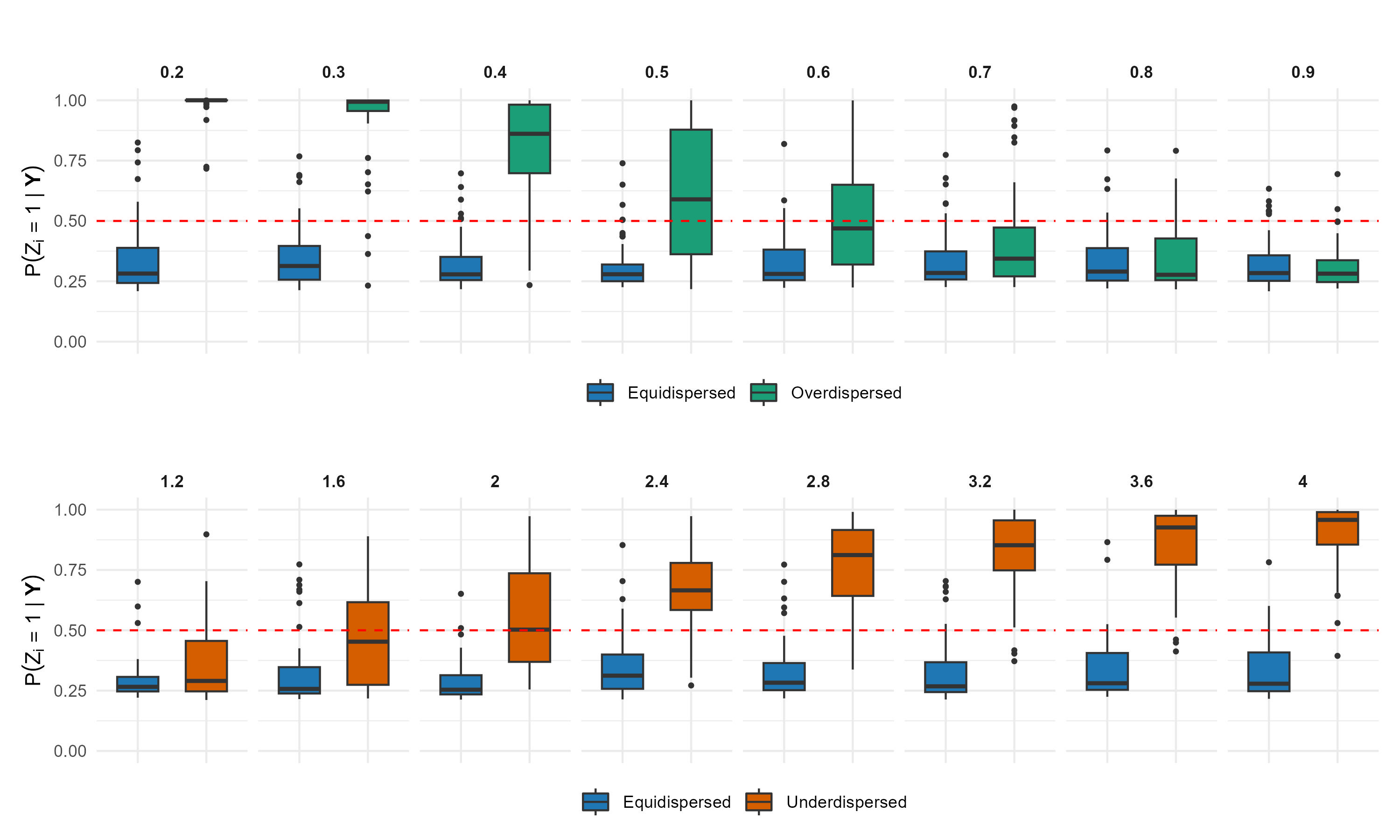}
\caption{Boxplot distribution of the posterior indicator probabilities \(P(Z_i = 1\mid \bm{Y})\), under varying \(\nu_i^{true}\) (top of each sub-plot). Observations are split in the \textit{dispersion groups} of the data-generating mechanism. Threshold of \(0.5\) is highlighted in a red dashed line.}
\label{fig:OD_UD_combined_sim}
\end{figure}

Figure \ref{fig:OD_UD_combined_sim} summarises the distribution of posterior probabilities \(P(Z_i = 1 \mid \bm{Y})\), for each set of simulations under varying \(k\). Each boxplot is formed by \(50\) posterior means of the respective latent indicators \(Z_i\), and are split according to the \textit{true} dispersion of the data-generating process. For each value of \(k\), the equidispersed groups are shown as a baseline or \textit{control} group. 
We can see from the plot that the posterior probabilities of \(Z_i\) for the equidispersed teams are concentrated around a baseline of \(\approx 0.25\). This value reflects our flat prior choices on \(p_i\) and \(\eta_i\), the variability arising from small sample sizes, and the fact that the slab is centred around the value of the spike distribution, all which contribute in preventing the posterior from collapsing completely to zero even for equidispersed data. For the non-equidispersed distributions, the behaviour of \(Z_i\) as a function of \(k\) is quite clear: we obtain almost perfect recovery (\(P(Z_i = 1) \approx 1\)) as \(\nu_{\text{true}}\) departs from \(1\), whereas they converge to the baseline as \(\nu_{\text{true}} \rightarrow 1\). Note that the indicators do not converge to \(1\) in the underdispersed examples since we are considering only a smaller subset of the parameter space, which theoretically extends up to \(\infty\).

Looking at a second diagnostic tool of our simulation study, we evaluate the model's ability to correctly identify dispersed units based on their posterior estimates. Specifically, we classify unit \textit{i} as dispersed whenever \(\mathbb{E}[P(Z_i = 1 \mid \bm{Y})] > 0.5\), i.e. the majority of the posterior samples select \(i\) to be CMP-distributed. This classification rule allows us to evaluate model performance in terms of \textit{true positives} (TP), when dispersed units are correctly identified (i.e., their indicator probability exceeds the threshold of \(0.5\)), and \textit{false positives} (FP), when equidispersed units are incorrectly classified as dispersed under the same criterion. 
Table \ref{tab:disp_sim_tp_fp} summarises the TP and FP over the varying values of \(k\). We expect the FP rate to be relatively stable across the simulations, given that it should not be influenced by the value of \(k\) from which the \textit{dispersed} data are generated with. Indeed, there does not seem to be big structural deviations from the average of \(\approx 10\%\). The presence of these false positives is not necessarily alarming either, as they naturally arise from the randomness of the data generation process over finite sample sizes. Finally, the TP classification is almost perfect for extremely over-dispersed settings \((\nu_i^{\text{true}} \leq 0.3)\), and performs very well up to \(\nu_i^{\text{true}} \leq 0.6\), after which we observe a sharp decline in TP rate. This is mirrored for the under-dispersed simulations, where the classification is extremely good for \((\nu_i^{\text{true}} \geq 3.2)\), whereas the classification precision is greatly reduced at the value of \(\nu_i^{\text{true}} = 1.6\).
\begin{table}[htbp]
\centering
\small
\begin{tabular}{lcccccccccccccccc}
$\nu_i^{\text{true}}$ 
& 0.2 & 0.3 & 0.4 & 0.5 & 0.6 & 0.7 & 0.8 & 0.9 
& 1.2 & 1.6 & 2.0 & 2.4 & 2.8 & 3.2 & 3.6 & 4.0 \\
\midrule
\textbf{FP} 
& 0.12 & 0.16 & 0.12 & 0.10 & 0.10 & 0.16 & 0.12 & 0.16 & 0.06 & 0.14 & 0.04 & 0.12 & 0.10 & 0.12 & 0.12 & 0.06\\
\textbf{TP} 
& 1.00 & 0.94 & 0.90 & 0.58 & 0.48 & 0.22 & 0.14 & 0.04 & 0.18 & 0.42 & 0.52 & 0.84 & 0.84 & 0.94 & 0.94 & 0.98\\
\end{tabular}
\caption{True Positive (TP) and False Positive (FP) rates of dispersion classification across varying $\nu_i^{\text{true}}$.}
\label{tab:disp_sim_tp_fp}
\end{table}%

\subsection{Model Fit}
Aside from dispersion recovery, we can assess and compare overall model fit capabilities of the CMP model with respect to the Poisson baseline. In particular, we want to evaluate the performance of both the proposed CMP-SAS model and the CMP-Full model against the Poisson model, under three simulation settings of \textit{extreme}, \textit{medium}, and \textit{minimal} levels of non-equidispersion. 
These settings are chosen to be representative of varying data-generating scenarios, in order to showcase differences in model fit between the three models. For each of the settings, we generate three initialisations with different random seeds, and compute/estimate the WAIC according to Eq.[\ref{eq:waic}], and summarise the model fit results in Table \ref{tab:WAIC_Sim}.

In the extremely dispersed settings \((\nu_i^{\text{true}} \in \{0.3, 4\})\), expectedly, both CMP models vastly outperform the Poisson baseline, with minimal differences between the two. On the other hand, in the least dispersed simulations \((\nu_i^{\text{true}} \in \{0.9, 1.2\})\), the Poisson likelihood is shown to be the most appropriate, although the CMP-SAS model maintains very close performances in all simulations, displaying the robustness of the spike-and-slab specification to equidispersed settings, whereas the fully CMP model shows a more noticeable decay in WAIC. Lastly, for the mildly dispersed settings \((\nu_i^{\text{true}} \in \{0.6, 2\})\), the CMP-SAS model outperforms the other two models in consideration under all the replications of the simulations, showcasing a consistent improvement in model fit with respect to the other two models. 

Overall, the SAS specification is able to capture the presence of non-equidispersion, while maintaining robustness in potentially ambiguous cases of equidispersed data. Our simulations show that the CMP-SAS is more flexible than both the Poisson and CMP, as it provides competitive results when the model is misspecified, while excelling on the more challenging settings of mildly dispersed data.
\begin{table}[htbp]
    \centering
    \resizebox{\textwidth}{!}{
\begin{tabular}{lccccccccc}
\starttabularbody
\multicolumn{1}{c}{\(\bm{\nu_i^{\text{true}}}\)} & \multicolumn{3}{c}{\textbf{0.3}} & \multicolumn{3}{c}{\textbf{0.6}} & \multicolumn{3}{c}{\textbf{0.9}} \\
\cmidrule(l{3pt}r{3pt}){1-1} \cmidrule(l{3pt}r{3pt}){2-4} \cmidrule(l{3pt}r{3pt}){5-7} \cmidrule(l{3pt}r{3pt}){8-10}
Replicate & 1 & 2 & 3 &  1 & 2 & 3  & 1 & 2 & 3 \\
\cmidrule(l{3pt}r{3pt}){1-1} \cmidrule(l{3pt}r{3pt}){2-4} \cmidrule(l{3pt}r{3pt}){5-7} \cmidrule(l{3pt}r{3pt}){8-10}
\textbf{Poisson} & 2836.5 & 2789.2 & 2873.0 & 2263.9 & 2393.0 & 2354.2 & 2131.3 & \textbf{2189.9} & \textbf{2157.1} \\
\textbf{CMP-SAS} & \textbf{2640.5} & 2619.6 & \textbf{2656.6} & \textbf{2242.3} & \textbf{2374.9} & \textbf{2325.9} & \textbf{2130.5} & 2190.8 & 2159.9 \\
\textbf{CMP-Full} & 2646.4 &\textbf{ 2615.3} & 2657.7 & 2244.3 & 2375.2 & 2329.3 & 2133.7 & 2208.8 & 2176.0 \\
\midrule
\multicolumn{1}{c}{\(\bm{\nu_i^{\text{true}}}\)} & \multicolumn{3}{c}{\textbf{4}} & \multicolumn{3}{c}{\textbf{2}} & \multicolumn{3}{c}{\textbf{1.2}} \\
\cmidrule(l{3pt}r{3pt}){1-1} \cmidrule(l{3pt}r{3pt}){2-4} \cmidrule(l{3pt}r{3pt}){5-7} \cmidrule(l{3pt}r{3pt}){8-10}
Replicate & 1 & 2 & 3 &  1 & 2 & 3  & 1 & 2 & 3 \\
\cmidrule(l{3pt}r{3pt}){1-1} \cmidrule(l{3pt}r{3pt}){2-4} \cmidrule(l{3pt}r{3pt}){5-7} \cmidrule(l{3pt}r{3pt}){8-10}
\textbf{Poisson} & 1838.5 & 1841.3 & 1861.5 & 1918.9 & 1914.9 & 1939.3 & \textbf{2060.3} & 2055.1 & \textbf{2109.8} \\
\textbf{CMP-SAS} & 1752.4 & 1754.5 & 1783.2 & \textbf{1914.2} &\textbf{1899.1} & \textbf{1931.0} & 2066.5 & \textbf{2054.3} & 2115.8 \\
\textbf{CMP-Full} &\textbf{1749.9} & \textbf{1751.5} & \textbf{1781.0} & 1920.3 & 1908.1 & 1937.3 & 2081.2 & 2067.8 & 2125.2 \\
\end{tabular}
}
    \caption{Computed WAIC over simulated data (\(n = 760\) for each simulated league), under different underlying generating true dispersion \(\nu_i^{\text{true}}\). Generating dispersion are chosen to represent extreme, mild, and negligible non-equidispersion. For each value of dispersion, three different random replications are shown. WAIC is computed for the three models under consideration, where the best model's result for each simulation is highlighted in bold.}
    \label{tab:WAIC_Sim}
\end{table}
\vspace{-1.5cm}
\section{Application}\label{ch:applications}
In this section, we illustrate the application of the CMP models to real football data from the English Premier League (EPL, or PL). In particular, we apply the methods to each season individually, as football teams drastically change in composition and technical staff between each season. To showcase the proposed model's properties and practical relevance, we first present the parameter inference results for the 2023/24 season in Section \ref{sec:par_inf}, the results on model fit and out-of-sample predictive performance in Section \ref{ch:model_eval}, and then extend the analysis on the last 5 seasons of the EPL in Section \ref{sec:more_seasons}.

\begin{figure}[htbp]
    \centering
    \includegraphics[width=0.8\linewidth]{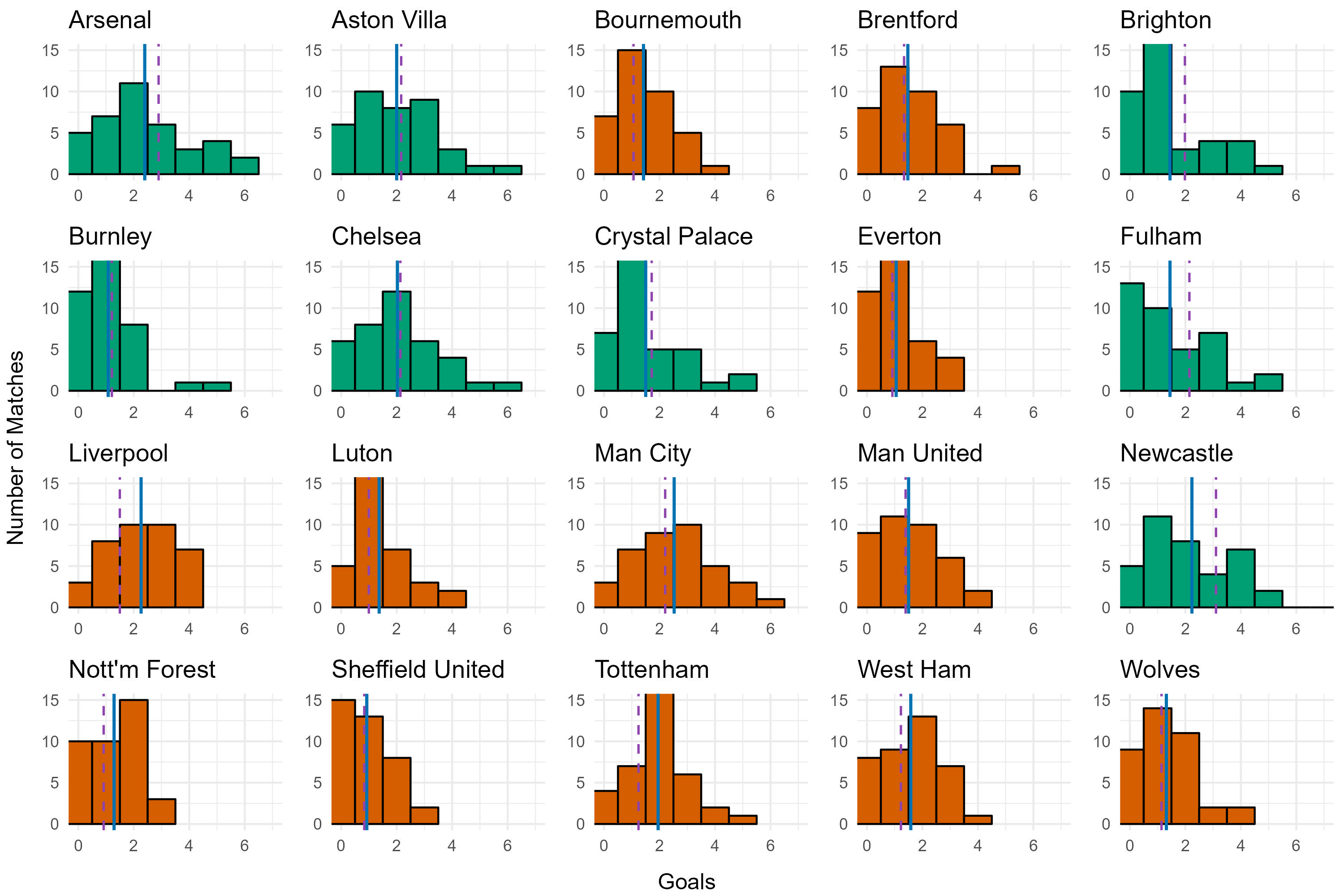}
    \caption{Aggregated histograms for the scores of each team in the Premier League season 2023/24. Solid blue and dashed purple lines represent the empirical mean and variance, respectively. The histograms are color coded in \textit{orange} to indicate empirical \textit{underdispersion} and in \textit{green} to denote empirical \textit{overdispersion}.}
    \label{fig:hist_individual_teams}
\end{figure}

As highlighted in Figure \ref{fig:hist_individual_teams}, the empirical distributions of the aggregate goal frequencies for each team across a season display differences in overall shape, reflecting varying degrees and directions of non-equidispersion. This serves as the primary motivation behind our modelling choice of tying the dispersion parameter to the scoring team, while also permitting additional inferential interpretation of this parameter with respect to team strategy.

The 2023/24 EPL season was characterised by a closely contested title race and a historically high scoring rate. Across all 380 matches, the season produced a record 1,246 goals, corresponding to an average of approximately 3.28 goals per match, the highest total the league ever registered. \textit{Manchester City} won the league, finishing marginally ahead of \textit{Arsenal} and \textit{Liverpool}, where these top-three teams distanced themselves considerably from the rest of the competition; at the lower end of the table, the bottom 3 teams in \textit{Luton}, \textit{Burnley} and \textit{Sheffield} performed very poorly both offensively and defensively, recording substantially lower goal differences than the other teams, see Appendix \ref{app:PL2324} for final results and statistics of the season. The high overall scoring rate and the pronounced heterogeneity in team-level goal statistics potentially suggest a departure from the equidispersion assumption: while most teams are relatively well-modelled by the Poisson likelihood, we find that 4 teams are significantly non-equidispersed according to the classification rule introduced in Section \ref{sec:sim_dispersion_recovery}. 

\vspace{-1cm}
\subsection{Parameters Inference and Interpretation}\label{sec:par_inf}
\begin{figure}[ht]
    \centering
    \includegraphics[width = 0.7\textwidth]{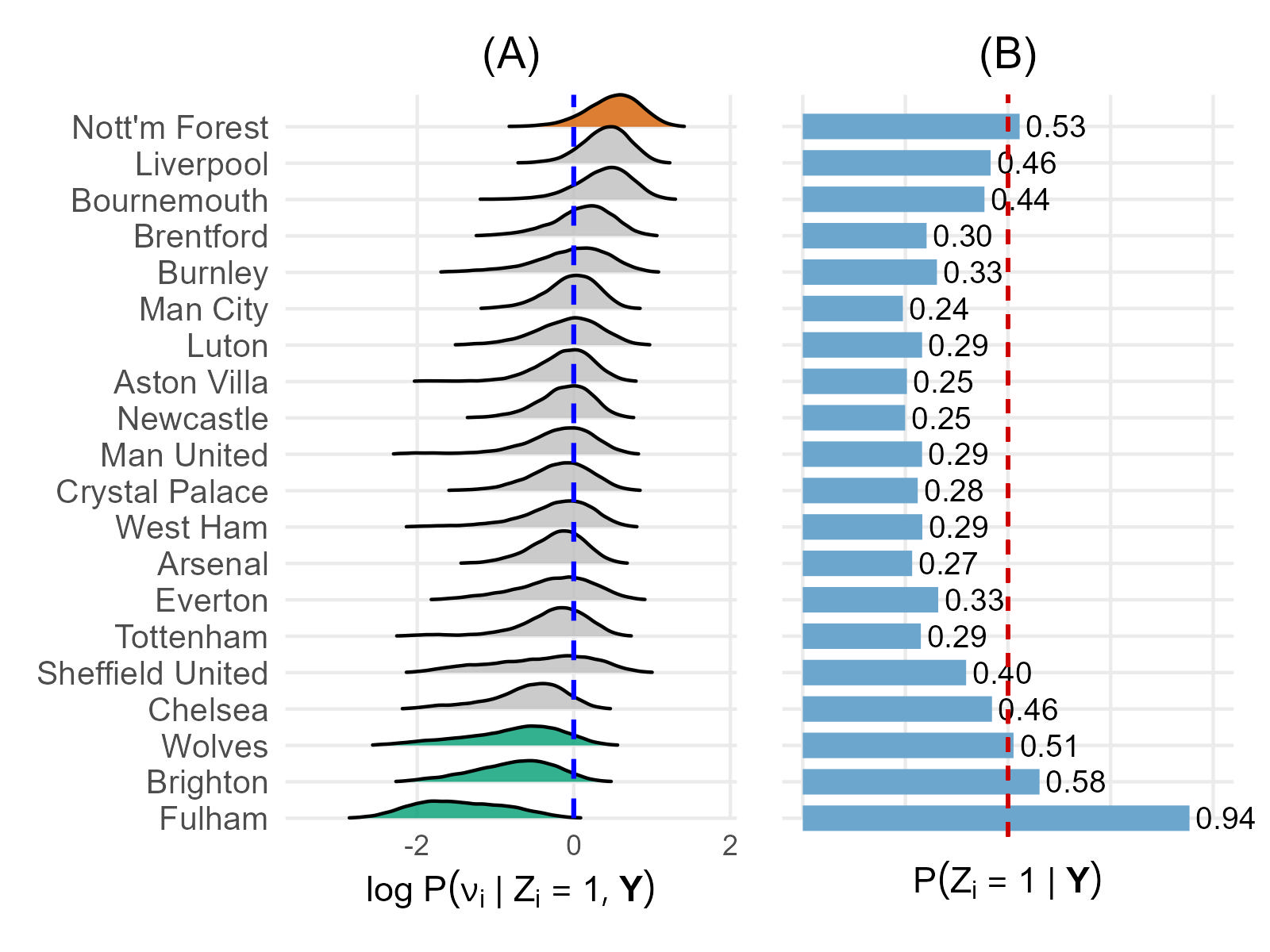}
    \caption{(A) Log-posterior dispersion distribution, conditional on \(Z_i = 1\), for the 20 teams of the 2023/24 season of the PL. (B) posterior \textit{slab} probability for each team \(i\). Densities are color-coded if \(P(Z_i = 1 \mid \bm{Y}) > 0.5\), in green (overdispersed) and orange (underdispersed).}
    \label{fig:Disp_joyplot_PL2324}
\end{figure}
We fit a standard Poisson as in \citet{Maher1982modelling}, our proposed CMP-SAS, and the slab-only CMP-Full model, using the MWGS algorithm detailed in Section \ref{sec:mwgb_algorithm}. Each model is run for \(250,000\) iterations, conservatively discarding \(50,000\) samples as burn-in. We begin the posterior analysis by examining the dispersion parameters, which constitute the primary innovation of the CMP-SAS model with respect to the Poisson distribution. Figure \ref{fig:Disp_joyplot_PL2324} summarises the results for the dispersion in the SAS model. Specifically, sub-plot (A) shows the posterior log-distribution of the dispersion coefficient \(\nu_i\) for each team \(i\), conditional on the indicator \(Z_i = 1\), corresponding to the \textit{slab} part of the distribution. Sub-plot (B) shows the posterior \textit{slab} probability \(P(Z_i = 1 \mid \bm{Y})\), representing the probability that team \(i\) follows a CMP distribution rather than a Poisson distribution. Values approaching 1 indicate strong evidence in favor of the CMP distribution, whereas values toward 0 suggest that the baseline Poisson distribution is appropriate. Team goals consistent with a Poisson-generating process exhibit a posterior probability of approximately 0.25, as observed in the simulation studies (Section \ref{sec:sim_dispersion_recovery}). 

These probabilities allow for classification of teams based on their dispersion using a thresholding approach: for example, selecting a threshold of 0.5 would be an intuitive and natural choice, as it implies that more than 50\% of the posterior distribution for that team supports the CMP model. This threshold identifies 3 over-dispersed teams in \textit{Fulham, Brighton and Wolves}, and 1 under-dispersed team in \textit{Nottingham Forest} for the 2023/24 season of the PL. In practice, however, a slightly lower threshold of 0.4 may also be appropriate, as teams above this threshold still exhibit notable levels of non-equidispersion. With this lower threshold, three additional teams (\textit{Liverpool, Bournemouth and Chelsea}) would be included, each showing some evidence of non-equidispersion.

\begin{figure}[htbp]
    \centering
    \includegraphics[width=\linewidth]{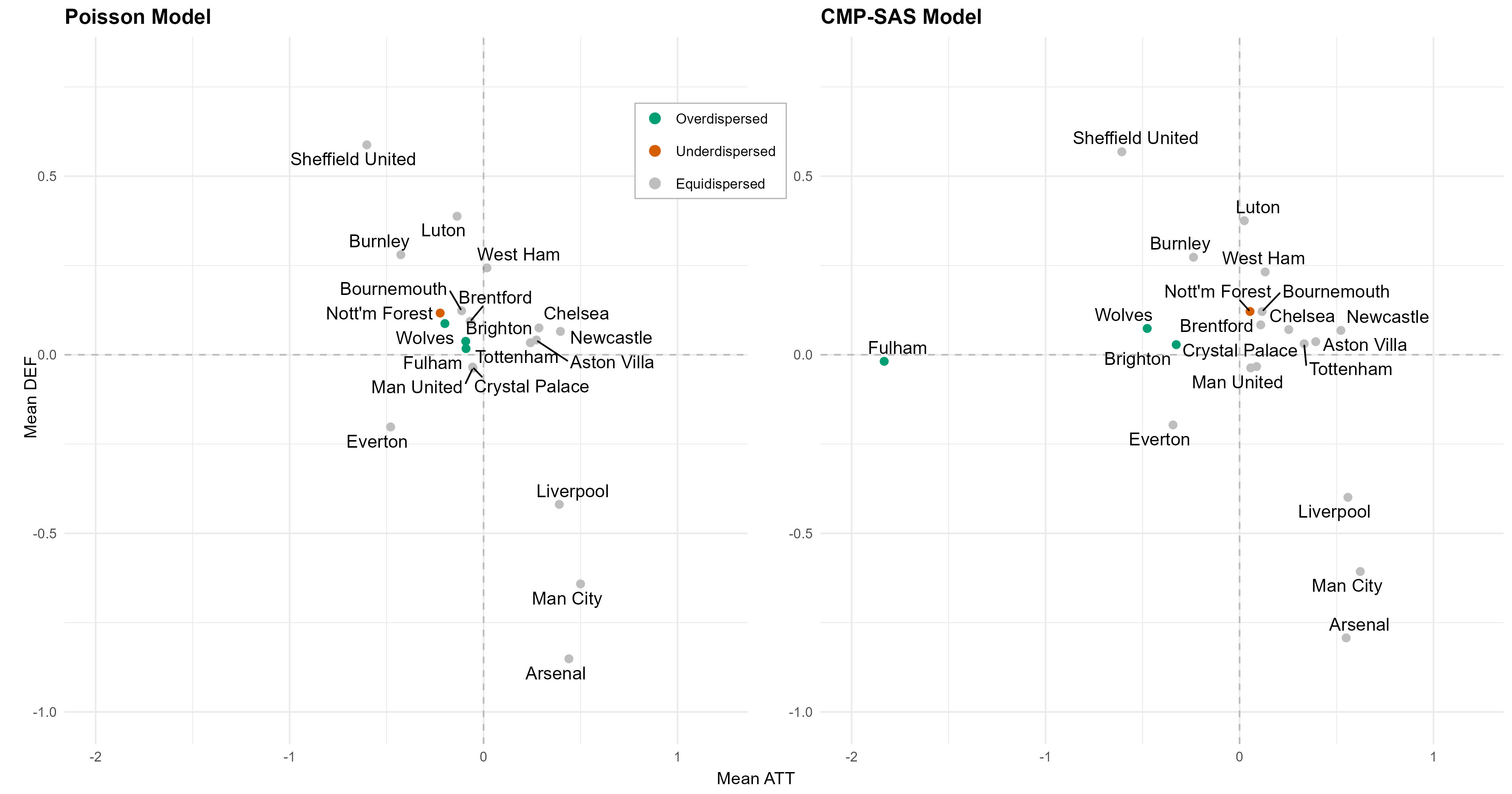}
    \caption{Median posterior estimates for the attack and defense parameters of the 20 teams of the Premier League season 2023/24, for the Poisson model (left) and the CMP-SAS model (right). Points are color-coded for teams \(i\) that have \(P(Z_i > 0.5 \mid \bm{Y})\) for ease of confrontation.}
    \label{fig:attdef_pl2324_comparison}
\end{figure}%
Figure \ref{fig:attdef_pl2324_comparison} provides a visual representation of the team-specific mean attack and defence parameters for the 20 teams of the season, where the origin represents the average of the league. A positive attack coefficient indicates above-average offensive capabilities, while a negative defence coefficient represents above-average defensive strength. A cluster of three teams formed by Manchester City, Arsenal, and Liverpool distance themselves from the rest in the bottom-right region of the plot, reflecting their status as the three dominant teams of the season. The addition of the dispersion parameter introduces some changes to the team-specific attack parameters, given its correlation with the dispersion in our formulation (Section \ref{sec:correlation_att_nu}). This difference can be quite noticeable for some of the teams, in particular to the ones that are pronouncedly non-equidispersed. The most extreme case is seen in \textit{Fulham}, which had a middle-of-the-pack mean attack parameter of approximately \(-0.1\) in the Poisson model, which turned to the worst attack in the league by far, with a value of approximately \(-1.8\) in the CMP-SAS model. This is caused by the extremely overdispersed parameter for the team, as highlighted from the previous figure, with a posterior mean of \(0.356\). The main interpretation of this behaviour is that \textit{Fulham} has a very high variance in their scoring behaviour, managing to score a high number of goals in many matches, but were not consistently doing so against the teams with weaker defence. Indeed, Figure~\ref{fig:hist_individual_teams} shows that this team exhibits a high frequency of scoreless matches, while also recording a substantial number of matches with three or more goals scored. A similar behaviour can be noted for the other \textit{overdispersed} teams, \textit{Brighton} and \textit{Wolves}, although to a lesser extent. Conversely, the \textit{underdispersed} teams have higher attack parameters with respect to the Poisson baseline, reflecting the tendency of these teams to score a consistent number of goals, closer to the mode of the distribution. In football, this trait could be explained by a \textit{conservative} way of playing to secure wins, once the team secures a lead. For instance, \textit{Liverpool} was one of the strongest teams in the league, averaging more than 2 goals per match, but ended up never scoring more than 4 goals throughout the entire season.

Table \ref{tab:pl2324_pars} provides an overview of the posterior means for all the parameters of interest for the Poisson baseline and the proposed CMP-SAS model. Overall, the point estimates of the defence parameters are highly similar across models, which is expected given that our defence specification is unchanged from the Poisson model. Consistently with the findings discussed above, the attack parameter estimates under the CMP–SAS model can exhibit substantial differences relative to the Poisson baseline. Moreover, their interpretation can no longer be made in isolation, but must instead be considered jointly with the corresponding dispersion parameters. In general, teams exhibiting over-dispersion have lower estimated attack parameters relative to the Poisson model, whereas for under-dispersed teams, they display higher attack estimates. 

{\setlength{\tabcolsep}{22pt}
\begin{table}[h!]
    \centering
    \footnotesize
    \resizebox{\textwidth}{!}{%
    \begin{tabular}{
                        l
                        S[table-format=1.3]
                        S[table-format=1.3]
                        S[table-format=1.3]
                        S[table-format=1.3]
                        S[table-format=2.3]
                        S[table-format=1.2]
                        }
\starttabularbody
\multicolumn{1}{l}{} & \multicolumn{2}{c}{\textbf{Poisson}} & \multicolumn{4}{c}{\textbf{CMP-SAS}} \\
\cmidrule(l{3pt}r{5pt}){1-1} \cmidrule(l{3pt}r{-8pt}){2-3} \cmidrule(l{15pt}r{3pt}){4-7}
\textit{Team} & {\(\qquad \ \ \alpha_i^{\text{Pois}}\)} & {\(\qquad \ \ \beta_i^{\text{Pois}}\)} & {\(\qquad \ \ \alpha_i^{\text{SAS}}\)} & {\(\qquad \ \ \beta_i^{\text{SAS}}\)} & {\(\qquad \ \ \nu_i\)} & {\(Z_i = 1\)}\\
\cmidrule(l{3pt}r{5pt}){1-1} \cmidrule(l{3pt}r{-8pt}){2-3} \cmidrule(l{15pt}r{3pt}){4-7}
\textit{Nott'm Forest} & -0.223 {(0.150)} & 0.117 {(0.127)} & 0.054 {(0.182)} & 0.121 {(0.127)} & 1.395 {(0.572)} & 0.53\\
\textit{Liverpool} & 0.390 {(0.110)} & -0.420 {(0.168)} & 0.559 {(0.123)} & -0.399 {(0.165)} & 1.261 {(0.423)} & 0.46\\
\textit{Bournemouth} & -0.113 {(0.141)} & 0.123 {(0.125)} & 0.116 {(0.189)} & 0.121 {(0.125)} & 1.241 {(0.447)} & 0.44\\
\textit{Brentford} & -0.069 {(0.140)} & 0.093 {(0.129)} & 0.110 {(0.177)} & 0.084 {(0.126)} & 1.059 {(0.255)} & 0.30\\
\textit{Burnley} & -0.426 {(0.167)} & 0.280 {(0.118)} & -0.237 {(0.304)} & 0.273 {(0.114)} & 1.025 {(0.275)} & 0.33\\
\textit{Man City} & 0.500 {(0.106)} & -0.642 {(0.185)} & 0.624 {(0.126)} & -0.607 {(0.178)} & 1.005 {(0.161)} & 0.24\\
\textit{Luton} & -0.137 {(0.147)} & 0.388 {(0.113)} & 0.025 {(0.214)} & 0.375 {(0.110)} & 0.999 {(0.214)} & 0.29\\
\textit{Newcastle} & 0.396 {(0.112)} & 0.065 {(0.133)} & 0.522 {(0.140)} & 0.068 {(0.130)} & 0.986 {(0.157)} & 0.25\\
\textit{Aston Villa} & 0.272 {(0.119)} & 0.041 {(0.135)} & 0.393 {(0.203)} & 0.037 {(0.130)} & 0.985 {(0.168)} & 0.25\\
\textit{Crystal Palace} & -0.056 {(0.140)} & -0.035 {(0.138)} & 0.089 {(0.205)} & -0.033 {(0.132)} & 0.973 {(0.189)} & 0.28\\
\textit{Man United} & -0.055 {(0.139)} & -0.037 {(0.137)} & 0.059 {(0.374)} & -0.037 {(0.134)} & 0.968 {(0.208)} & 0.29\\
\textit{West Ham} & 0.017 {(0.134)} & 0.243 {(0.121)} & 0.132 {(0.311)} & 0.232 {(0.118)} & 0.963 {(0.207)} & 0.29\\
\textit{Arsenal} & 0.440 {(0.107)} & -0.852 {(0.205)} & 0.549 {(0.149)} & -0.793 {(0.200)} & 0.962 {(0.161)} & 0.27\\
\textit{Everton} & -0.479 {(0.169)} & -0.203 {(0.148)} & -0.343 {(0.384)} & -0.196 {(0.142)} & 0.958 {(0.242)} & 0.33\\
\textit{Tottenham} & 0.241 {(0.120)} & 0.035 {(0.132)} & 0.333 {(0.300)} & 0.031 {(0.134)} & 0.952 {(0.195)} & 0.29\\
\textit{Sheffield United} & -0.602 {(0.186)} & 0.589 {(0.104)} & -0.606 {(0.736)} & 0.568 {(0.102)} & 0.919 {(0.318)} & 0.40\\
\textit{Chelsea} & 0.285 {(0.118)} & 0.074 {(0.133)} & 0.252 {(0.424)} & 0.070 {(0.130)} & 0.819 {(0.265)} & 0.46\\
\textit{Wolves} & -0.200 {(0.148)} & 0.087 {(0.129)} & -0.475 {(0.948)} & 0.074 {(0.127)} & 0.766 {(0.314)} & 0.51\\
\textit{Brighton} & -0.092 {(0.141)} & 0.037 {(0.130)} & -0.325 {(0.681)} & 0.028 {(0.131)} & 0.729 {(0.305)} & 0.58\\
\textit{Fulham} & -0.090 {(0.142)} & 0.017 {(0.134)} & -1.831 {(1.501)} & -0.018 {(0.130)} & 0.323 {(0.239)} & 0.94\\
\end{tabular}
}
\par\vspace{1em}\par
\centering
\begin{tabular}{S[table-format=1.3]
                S[table-format=1.3]}

\hspace{1cm}\text{Home-}\textbf{Poisson} & \hspace{1cm}\text{Home-}\textbf{CMP-SAS} \\
\midrule
 0.474 {(0.041)} & 0.379 {(0.065)}\\
\end{tabular}
    \caption{Posterior estimated means (standard deviation) of the parameters for the 2023/24 season of the Premier League, for the Poisson and the CMP-SAS models.}
    \label{tab:pl2324_pars}
\end{table}}
\clearpage
\subsection{Model Evaluation}\label{ch:model_eval}
Aside from inference, one of the key goals of football modelling is \textit{prediction}, which will be the focus of the rest of this chapter. In the following, we are interested in evaluating the better fit of the CMP likelihood, but also compare the usefulness of the spike-and-slab construction by including a fully CMP model in the comparison. We carry out evaluations both in-sample, by using all the data in a given season, and out-of-sample, by using only partial data and predicting future games.
\subsubsection{In-sample}
Using all the data in each season, we infer the parameters and compute the WAIC Eq.[\ref{eq:waic}] using \(S = 5,000\) samples from the posterior for each model. Table \ref{tab:WAIC_PL} shows the computed WAIC for season 2023/24 of the PL, together with their decomposition into the expected log predictive density and the effective complexity penalty. According to the WAIC, the CMP-SAS performs overall better, followed by the fully-CMP model, and lastly the Poisson baseline. In particular, the CMP–SAS model achieves higher log-predictive density than both alternatives, at the cost of only a moderate increase in model complexity.
\begin{table}[htbp]
    \centering
\begin{tabular}{lccc}
Model & lppd & \(\textbf{p}_{\text{WAIC}}\) & WAIC\\
\midrule
Poisson & -1159.8  & 40.4  & 2400.3\\
CMP-SAS & -1141.7 & 43.3 & \textbf{2370.0}\\
CMP-Full & -1143.7 & 49.3 & 2386.1\\
\end{tabular}
    \caption{WAIC breakdown for in-sample models computed on data from the Premier League season 23/24.}
    \label{tab:WAIC_breakdown_PL2324}
\end{table}
\vspace{-0.5cm}
\subsubsection{Out-of-sample}\label{sec:model_eval}
\begin{figure}[htbp]
    \centering
    \includegraphics[width=1\linewidth]{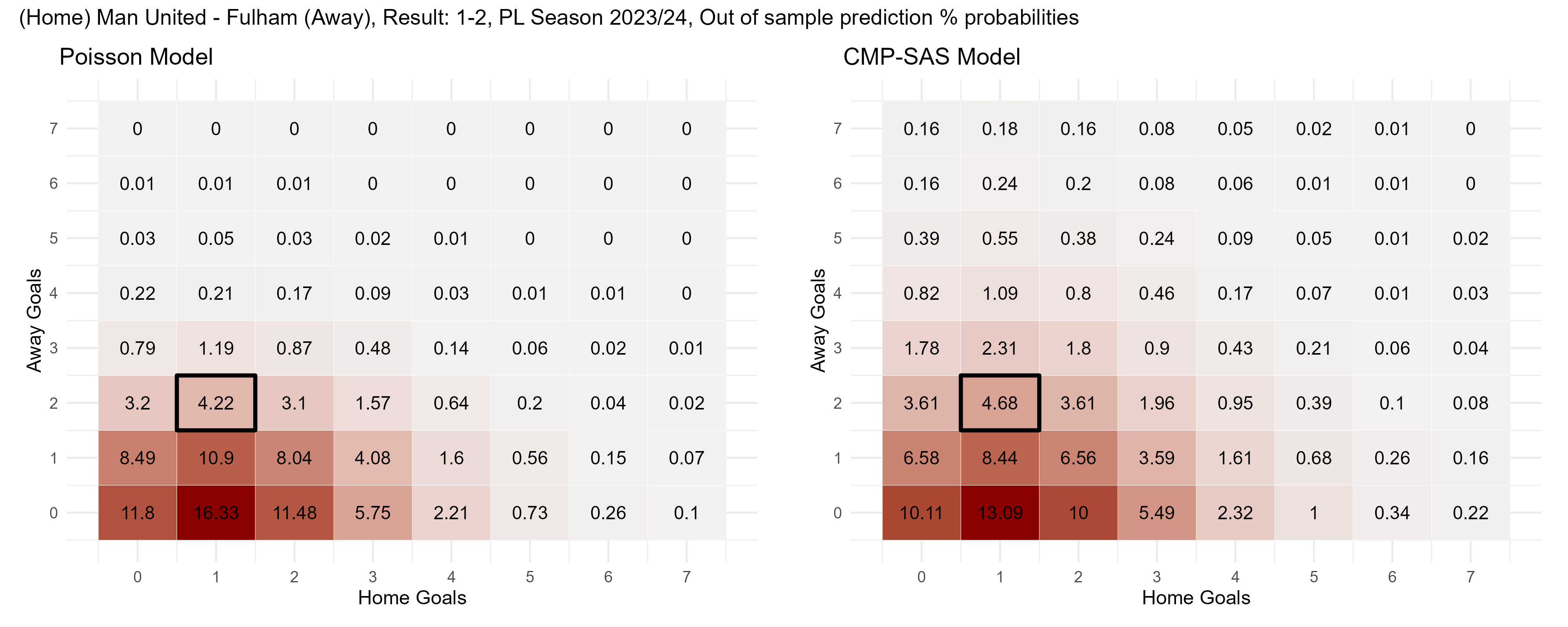}
    \caption{Exact result probabilities for each outcome from the Poisson (left) and CMP-SAS (right) model posterior simulation of the match between Manchester United and Fulham in the PL 2023/24 season. Observed outcome of 1-2 highlighted.}
    \label{fig:ful_tot_predictive}
\end{figure}

As previewed in Section \ref{ch:posterior_predictive}, we compute the out-of-sample predictive forecasts by estimating the parameters using all the matches up to that \textit{game-day}.
Figure \ref{fig:ful_tot_predictive} provides an example of the posterior probabilities of a particular match of the EPL season 2023/24 between \textit{Manchester United} and \textit{Fulham}. In this example, the CMP-SAS model assigns a slightly higher probability of \(4.68\%\) to the actual result of \(1-2\) with respect to the Poisson model, which predicts a probability of \(4.22\%\).
From these, it is also straightforward to obtain the probabilities of the possible outcomes of the matches of \textit{home win}, \textit{draw}, and \textit{away win}, which we denote as \(\{p^H, p^D, p^A\}\), respectively. Indeed, the anti-diagonal of the matrix sums to \(p^D\), and it splits the matrix into two triangular matrices which sum to \(p^H\) and \(p^A\). In the given example, the estimates for \(\{p^H, p^D, p^A\}\) are \(\{0.541, 0.263, 0.196\}\) for the Poisson model and \(\{0.497, 0.233, 0.270\}\) for the CMP-SAS model, a difference of about \(8\%\) of predicting the correct result of an away win. This large difference between the two predictions is given by the high degree of over-dispersion in Fulham's goals.

Furthermore, we aim to exploit the goal-count forecasts in greater detail. To this end, for each match we consider predictions for three types of outcomes:
 \begin{itemize}
     \item \textbf{Match Outcome} - the usual \textit{three-way} forecast of the final result of a match, \{home win, draw, away win\};
     \item \textbf{Over-Under} - which represents the \textit{two-way} final tally of total goals being \textit{over} or \textit{under} a certain threshold of goals, the most common of which is 2.5;
     \item \textbf{Goal-Difference} (GD) - which represents the difference in goals at the end of a match \(\text{GD} = \text{Home Goals} - \text{Away Goals}\), i.e a positive GD indicates a \textit{home win}, while a negative GD indicates an \textit{away win}, and 0 indicates a \textit{draw}.
 \end{itemize}
 These three types of result forecasting are also among the most popular categories in betting markets. In particular, Over-Under and Goal-Difference are particularly relevant for the CMP models as these are specifically designed to better capture the number of goals in a match, and not just the final result.
 
 Table \ref{tab:OOS_Eval_PL2324} contains the out-of-sample average IGN scores for the 3 models under consideration, for the season 2023/24 of the Premier League. Under all three markets in consideration, the CMP-SAS model is performing better than the Poisson baseline, with the most notable differences in the Over-Under 2.5 forecasts. In this case, the fully CMP model is performing worse than the Poisson baseline in both Outcome and GD forecasts.
\begin{table}[htbp]
    \centering
\begin{tabular}{lccc}
\textbf{IGN} & Outcome & Over-Under 2.5 & Goal Difference\\
\midrule
Poisson & 1.327 & 0.974 & 2.940\\
CMP-SAS & \textbf{1.326} & 0.929 & \textbf{2.937}\\
CMP-Full & 1.330 & \textbf{0.925} & 2.948\\
\end{tabular}
    \caption{Average Ignorance Score  over second half of the league \((n = 190)\), for the out-of-sample forecasts for the three models in the PL season 2023/24. Lowest score is in bold.}
    \label{tab:OOS_Eval_PL2324}
\end{table}
\subsection{Analysis of 5 Seasons}\label{sec:more_seasons}
\begin{figure}[htbp]
    \centering
\includegraphics[width=1\linewidth]{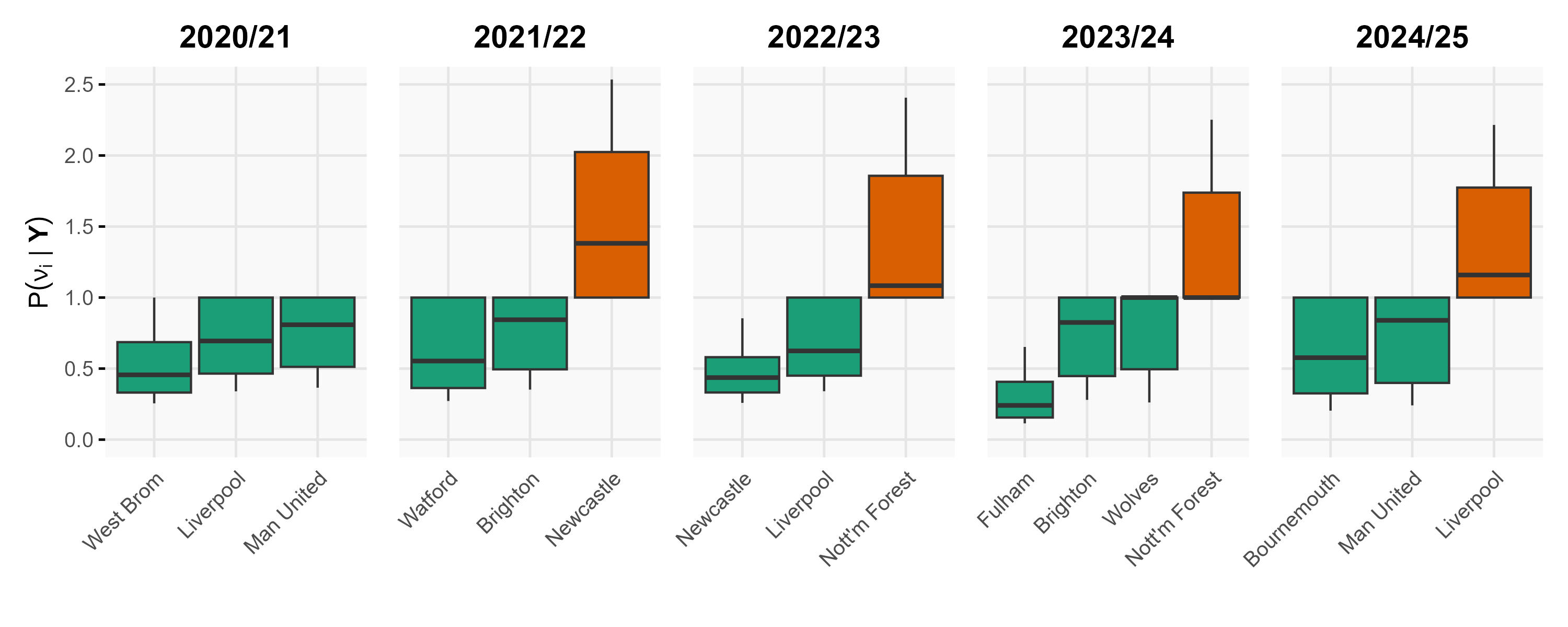}
\caption{Posterior dispersions boxplots \(P(\nu_i \mid \bm{Y})\) for the teams such that \(P(Z_i = 1 \mid \bm{Y}) \geq 0.5\), for the 5 seasons of the Premier League from 2020/21 up to 2024/25.}
\label{fig:pl_5seas_disp_boxplots}
\end{figure}

We now extend our analysis to the last 5 seasons of the Premier League. Figure \ref{fig:pl_5seas_disp_boxplots} shows the log-dispersion posterior summaries for the teams that go above the threshold \(P(Z_i \mid \bm{Y}) > 0.5\) for each season. Over the time period considered, there were 3 dispersed teams per season, with the exception of season 2023/24 which contained 4. The majority of them display \textit{overdispersion}, which is perhaps consistent with the EPL being one of the football leagues with the highest average number of goals. Liverpool is represented in 3 out of 5 years as non-equidispersed, albeit changing in the direction of dispersion, possibly indicating changes of playstyle or strategy between these seasons. The rest of the teams do not seem to show continuity in non-equidispersion across seasons, with the exception of \textit{Nottingham Forest}, which is \textit{underdispersed} for two years in a row, from season 2022/23 to 2023/24.

Turning to the model comparison, the WAIC provides clear evidence in favor of the CMP-SAS specification. As shown in Table \ref{tab:WAIC_PL}, the CMP-SAS model achieves improved fit across all seasons in the period considered. This consistently lower WAIC indicates that Premier League scoring data exhibit meaningful departures from equidispersion, thereby motivating the use of CMP-based models over the conventional Poisson formulation.
\begin{table}[bhp]
    \centering%
    \caption{WAIC for in-sample models computed on data from the Premier League seasons from 2020/21 to 2024/25.}
\begin{tabular}{lccccc}
\textbf{WAIC} & 2020/21 & 2021/22 & 2022/23 & 2023/24 & 2024/25 \\
\midrule
Poisson & 2296.4 & 2242.2 & 2286.1 & 2400.3 & 2312.8 \\
CMP-SAS &  \textbf{2280.5} & \textbf{2236.4} & \textbf{2277.9} & \textbf{2370.0} & \textbf{2293.1} \\
CMP-Full & 2290.1 & 2246.2 & 2287.5 & 2386.1 & 2305.6 \\

\end{tabular}
    \label{tab:WAIC_PL}
\end{table}

We next evaluate out-of-sample predictive performance using the average Ignorance Scores (IGN) Eq. [\ref{eq:ignorance_score}] computed on the second half of each of the five seasons under study (190 games). Results are summarised in Table \ref{tab:oos_pl_5seasons_ign}, where overall, the CMP-based models deliver modest but systematic improvements over the Poisson benchmark. Although the magnitude of these improvements varies across seasons and forecast types, the CMP-SAS model consistently outperforms the Poisson model in every season considered, suggesting that the gains in within-sample fit translate effectively to out-of-sample predictive performance, with no clear evidence of substantial overfitting. In contrast, the fully CMP specification does not exhibit the same level of consistency: it underperforms relative to the Poisson benchmark in some seasons and forecast markets, while in cases where it performs better, the gains are generally comparable to those achieved by the CMP-SAS model. These results further support the value of the SAS formulation, preserving the flexibility afforded by the CMP likelihood, while offering benefits in terms of parameter inference, interpretability, and robustness to equidispersion.

\begin{table}[htbp]
    \centering
    \begin{tabular}{lccccc}

\textbf{Season}  &  2020/21 & 2021/22 & 2022/23 & 2023/24 & 2024/25 \\
\midrule
\textbf{Match Outcome} & & & & &\\
\midrule
Pois & 1.497 & 1.403 & \textbf{1.401} & 1.327 & 1.415\\
CMP-SAS  & \textbf{1.480} & 1.399 & \textbf{1.401} & \textbf{1.326} & {1.397}\\
CMP-Full & {1.481} & \textbf{1.396} & 1.404 & 1.330 & \textbf{1.391}\\
\midrule
\textbf{Over-Under 2.5} & & & & &\\
\midrule
Pois & 1.019 & 1.063 & {0.961} & 0.974 & 1.025\\
CMP-SAS  & 0.998 & \textbf{1.056} & \textbf{0.960} & 0.929 & \textbf{1.004}\\
CMP-Full & \textbf{0.997} & 1.064 & 0.967 & \textbf{0.925} & {1.011}\\
\midrule
\textbf{Goal Difference} & & & & &\\
\midrule
Pois & 2.801 & 2.948 & 2.878 & 2.940 & 2.858\\
CMP-SAS & \textbf{2.796} & {2.936} & \textbf{2.861} & \textbf{2.937} & \textbf{2.845}\\
CMP-Full & 2.806 & \textbf{2.932} & 2.866 & 2.948 & 2.846\\
\end{tabular}
    \caption{Out-of-sample average Ignorance scores for the second half of each league \((n = 190)\), for the seasons from 2020/21 up to 2024/25, for the three models under consideration and for three types of market forecasting.}
    \label{tab:oos_pl_5seasons_ign}
\end{table}
\FloatBarrier
\section{Discussion and Conclusions}\label{ch:conclusions}
In this work, we have implemented a generalisation of the \textit{conditionally independent} Poisson goal model \citep{Maher1982modelling} through the Conway-Maxwell-Poisson likelihood. In particular, we include a flexible spike-and-slab prior specification over the dispersion parameters, capturing simultaneously the probability of departure from equidispersion and the level of dispersion in goal-scoring for each team. Model inference is carried out in a Bayesian framework through a Metropolis-within-Gibbs sampler, leveraging the methods developed by \citet{Benson21} to overcome the intractability of the CMP likelihood. 
The main contribution of this novel framework lies in the ability to correctly separate heterogeneities in the dispersion behaviour of the data. Indeed, we show that it is effective in capturing both over-dispersion and under-dispersion, while remaining robust to equidispersed settings.  Dispersion retrieval capabilities are confirmed through an extensive set of simulation studies under varying levels of non-equidispersion. The methodology is then applied to real data from the English Premier League, showcasing its practical applications through new interpretations of latent characteristics behind goal-scoring behaviour of football teams. In particular, we show that some teams can display high variance in their offensive capabilities, explaining their inconsistent scoring patterns throughout a given season. Conversely, other teams are found to be under-dispersed, with the interpretation that they record a relatively stable number of goals, close to their average. It is worthy to note that these patterns are not directly related to the overall skill of each team, as non-equidispersion occurs across teams with diverse final standings within each season.

The proposed model complements the existing literature on dispersed count data, by demonstrating how the CMP likelihood with SAS priors can be leveraged for interpretable, unit-specific inference on non-equidispersion. Our results show that explicitly accounting for dispersion leads to  effective and consistent improvements for in-sample model fit and predictive performance in Premier League scoring data with respect to the Poisson baseline. It also generalises well to out-of-sample predictive tasks, out-performing the comparison models across three types of match-level forecasts. While these differences might not be substantial on average, they can be meaningful in the context of specific teams, where accounting for dispersion could generate more accurate odds for specific matches. This suggests potential applicability to betting contexts, although further refinements could be applied. In particular, the model under-predicts draws, which is likely attributable to the conditional independence assumption between the two match scores. Additionally, the current parameter estimation treats all historical matches equally, whereas a weighting scheme that emphasises recent results may better reflect temporal variation in team performance. Both of these enhancements are applied to the original Poisson framework by \citet{Dixon1997}, and are expected to improve predictive performance in the CMP setting as well.

The introduced CMP-SAS model offers an attractive balance between flexibility, robustness to equidispersion, and interpretability, enabling probabilistic identification of dispersion while retaining a clear connection to the underlying count data-generating process. Beyond football, the proposed framework provides a general and extensible approach for modelling count data in settings where both over- and under-dispersion may arise. This work illustrates how spike-and-slab constructions can be combined with the CMP likelihood to make principled, unit-level inferences on dispersion in complex and heterogeneous count data models.

\begin{repository}
    Historical data of football matches are obtained from \url{https://www.football-data.co.uk/}. All the code to reproduce the figures, tables and analyses in this paper are openly available at the following github repository: \url{https://github.com/nzhang98/compoisson_goal}
\end{repository}


\begin{funding}
 This publication has emanated from research conducted with the financial
support of Taighde Éireann – Research Ireland under Grant number $18/CRT/6049$. The Insight
Centre for Data Analytics is supported by Science Foundation Ireland under Grant Number
$12/RC/2289\_P2$.
\end{funding}


\printbibliography

\clearpage
\appendix

\section{- MCMC Strategy for the CMP likelihood \newline\citep{Benson21}}\label{app:A}
\subsection{Exchange MCMC}\label{app:exchange_mcmc}
This method, proposed by \citep{Murray12}, based on a previous work by \citep{Moller06}, is an ingenious solution that circumvents the need to evaluate the full likelihood of doubly-intractable posterior distributions. The main idea involves augmenting the posterior distribution with auxiliary data \(y'\), independent of \(y\), by changing the proposal distribution in the following convenient form:
\begin{align*}
    h(y', \bm{\theta}^*|y, \bm{\theta}) = h(y'|\bm{\theta}^*, \bm{\theta}, y) k(\bm{\theta}^*|y,\bm{\theta}) = \ &h(y'|\bm{\theta}^*) k(\bm{\theta}^*|\bm{\theta}), \\
    &h(y'|\bm{\theta}^*) \coloneqq \frac{q_f(y'|\bm{\theta}^*)}{\mathcal{Z}_f(\bm{\theta}^*)}.
\end{align*}
This causes the \textit{exchange} acceptance ratio of the augmented posterior to simplify in the following manner:
\begin{align*} \label{eq:mh_exch_acceptance_ratio}
    \alpha_{EX}(\bm{\theta},\bm{\theta}^*) &= \min{\left\{
    1, \frac{\frac{q_f(y|\bm{\theta}^*)}{\mathcal{Z}_f(\bm{\theta}^*)}}{\frac{q_f(y|\bm{\theta})}{\mathcal{Z}_f(\bm{\theta})}}\frac{h(y',\bm{\theta}|y, \bm{\theta}^*)}{h(y', \bm{\theta}^*|y, \bm{\theta})}\frac{p(\bm{\theta}^*)}{p(\bm{\theta})}
    \right\}} \nonumber\\
    &= \min{\left\{
    1, \frac{\frac{q_f(y|\bm{\theta}^*)}{\mathcal{Z}_f(\bm{\theta}^*)}}{\frac{q_f(y|\bm{\theta})}{\mathcal{Z}_f(\bm{\theta})}}\frac{\frac{q_f(y'|\bm{\theta})}{\mathcal{Z}_f(\bm{\theta})} k(\bm{\theta}|\bm{\theta}^*)}{\frac{q_f(y'|\bm{\theta}^*)}{\mathcal{Z}_f(\bm{\theta}^*)} k(\bm{\theta}^*|\bm{\theta})}\frac{p(\bm{\theta}^*)}{p(\bm{\theta})}
    \right\}} \nonumber\\
    & = \min{\left\{
    1, \frac{q_f(y|\bm{\theta}^*)}{q_f(y|\bm{\theta})}\frac{q_f(y'|\bm{\theta})} {q_f(y'|\bm{\theta}^*)}\frac{k(\bm{\theta}|\bm{\theta}^*)}{k(\bm{\theta}^*|\bm{\theta})}\frac{p(\bm{\theta}^*)}{p(\bm{\theta})}
    \right\}},  
\end{align*}
which is fully tractable, and crucially, the marginal for \(\bm{\theta}\) in the augmented posterior, under suitable conditions, is the \textit{original target} posterior distribution. Among these conditions is that we require \textit{exact} samples from the likelihood \(f(y|\cdot)\) at each proposed state \(\bm{\theta}, \bm{\theta}^*\). \citet{Benson21} describe the steps to obtain these samples through a rejection sampler.
\subsection{Rejection Sampler}\label{app:A_rej_sampler}
Denote the target density as:
\begin{equation*}
    f(y|\theta) = \frac{q_f(y|\theta)}{\mathcal{Z}_f(\theta)}.
\end{equation*}
To generate draws from \(f(y|\theta)\), we consider an envelope distribution \(g(y|\gamma)\) parametrised by \(\gamma \in \Gamma\). Assume that the envelope density can be written in similar form as:
\begin{equation*}
    g(y|\gamma) = \frac{q_g(y|\gamma)}{\mathcal{Z}_g(\gamma)}, \qquad \mathcal{Z}_g(\gamma) = \int_y q_g(y|\gamma) \ dx,
\end{equation*}
where the normalizing constant \(\mathcal{Z}_g(\gamma)\) may be tractable or intractable. Conditions on \(g\) are that it dominates the support of \(f\) and there is a positive enveloping constant \(M\) that satisfies \(Mg(y|\gamma) > f(y|\theta)\) for all \(y\). The optimal \(M\) is found as:
\begin{align}\label{eq:m_fg_equation}
    M_{f/g} = \sup_y\left\{\frac{f(y|\theta)}{g(y|\gamma)}\right\} = \frac{1/\mathcal{Z}_f(\theta)}{1/\mathcal{Z}_g(\gamma)}\sup_x\left\{\frac{q_f(y|\theta)}{q_g(y|\gamma)}\right\} = \frac{\mathcal{Z}_g(\gamma)}{\mathcal{Z}_f(\theta)}B_{f/g}.
\end{align}
The tractable acceptance probability of a sample \(y^*\) from the rejection sampler is then as follows:
\begin{equation}\label{eq:acceptance_rej}
    \alpha_{Rej}(y^*) = \frac{f(y^*|\theta)}{\left(\frac{\mathcal{Z}_g(\gamma)}{\mathcal{Z}_f(\theta)}B_{f/g}\right)g(y^*|\gamma)} = \frac{q_f(y^*|\theta)/\mathcal{Z}_f(\theta)}{\left(\frac{\mathcal{Z}_g(\gamma)}{\mathcal{Z}_f(\theta)}B_{f/g}\right)q_g(y^*|\gamma)/\mathcal{Z}_g(\gamma)} = \frac{q_f(y^*|\theta)}{B_{f/g}q_g(y^*|\gamma)}.
\end{equation}
For the COM-Poisson, following the method developed by \citep{Benson21}, we need to choose two envelope distributions for when \(\nu < 1\) (overdispersion) and for when \(\nu \geq 1\) (underdispersion). For the former case, we use a \textit{geometric envelope}, while for the latter case we employ a \textit{Poisson envelope}:
\begin{equation*}
    g(y|\gamma) = \begin{cases}
        g(y|\gamma = p) = p(1-p)^y, &\textit{if } \nu < 1, \\
        g(y|\gamma = \mu) = \frac{\mu^y}{e^\mu y!}, &\textit{if } \nu \geq 1,
    \end{cases} 
\end{equation*}
with associated tractable enveloping bounds:
\begin{equation}\label{eq:enveloping_bounds}
    B_{f/g} = \begin{cases}
        \frac{1}{p}\frac{\mu^{\left(\nu \floor*{\frac{\mu}{(1-p)^{1/\nu}}}\right)}}{(1-p)^{\left(\nu \floor*{\frac{\mu}{(1-p)^{1/\nu}}}\right)}\left(\nu \floor*{\frac{\mu}{(1-p)^{1/\nu}}}!\right)^\nu}, &\textit{if } \nu < 1, \\
        \left(\frac{\mu^{\floor*{\mu}}}{\floor*{\mu}!}\right)^{\nu-1}, &\textit{if } \nu \geq 1.
    \end{cases}
\end{equation}
Details about the justification and proof of the bounds for these two envelopes can be found in Theorem 3.1 of \citet{Benson21}. Although the rejection sampler is valid for any choice of \(p \in (0,1]\), the choice of the parameter \(p\) in the geometric envelope is made as to maximise the acceptance rate. Although the computation of the optimal \(p\) is not available in closed form, a good choice for it is chosen in \citet{Benson21} in order to match the first moment of the geometric distribution to the first moment approximations of the CMP distribution, obtaining:
\begin{equation*}\label{eq:p_geom_eq}
    \frac{1-p}{p} = \mu + \frac{1}{2\nu} - \frac{1}{2} \iff p = \frac{2\nu}{2\mu\nu + \nu + 1}
\end{equation*}
\begin{algorithm}[h]
\caption{Sampler for COM-Poisson(\(\mu, \nu\)) random variables}\label{alg:1_Exch_COMPois}
\KwIn{Parameters \(\theta = (\mu,\nu)\)}
START \\
\If{\(\nu \geq 1\)}{
    Sample \(y' \sim \text{Poisson}(\mu)\) \\
    Compute \(B_{f/g}^{[\nu\geq1]}\) using Eq.[\ref{eq:enveloping_bounds}] \\
    Set acceptance ratio as \(\alpha_{\text{rej}} = \frac{(\mu^{y'}/y'!)^\nu}{B_{f/g}^{[\nu\geq1]} \ (\mu^{y'}/y'!)}\) according to Eq.[\ref{eq:acceptance_rej}]\\
}
\If{\(\nu \leq 1\)}{
    Compute \(p = \frac{2\nu}{2\mu\nu + 1 + \nu}\) and sample \(y' \sim \text{Geometric}(p)\) \\
    Compute \(B_{f/g}^{[\nu\leq1]}\) using Eq.[\ref{eq:enveloping_bounds}] \\
    Set acceptance ratio as \(\alpha_{\text{rej}} = \frac{(\mu^{y'}/y'!)^\nu}{B_{f/g}^{[\nu\leq1]} \ p(1 - p)^{y'}}\) according to Eq.[\ref{eq:acceptance_rej}]\\
}
Generate \(u \sim \text{Uniform}(0,1)\) \\
\eIf{\(u \leq \alpha_{\text{rej}}\)}{
    \KwRet{\(y'\)} \\
}{
 GO TO START 
}
\end{algorithm}
\subsection{Unbiased estimator for intractable likelihood}\label{appendix:llhood_estimator}
Using Eq.[\ref{eq:m_fg_equation}], we can rewrite the normalizing constant of the likelihood as:
\begin{equation}\label{eq:link_normalizing_Zf}
    \frac{1}{\mathcal{Z}_f(\theta)} = \frac{1}{\mathcal{Z}_g(\gamma)}\frac{M_{f/g}}{B_{f/g}},
\end{equation}
where all the terms on the right-hand side are known (with the weak assumption that \(\mathcal{Z}_g(\gamma)\) is known) except the intractable bound \(M_{f/g}\), which can be estimated by running the rejection sampling at the given parameter \(\theta\) and record the number of draws \(n_r\) required to obtain \(r\) acceptances. The unbiased estimate for the bound based on \(r\) is given as:
\begin{equation*}
    \widehat{M}^{(r)}_{f/g} = \frac{n_r}{r}.
\end{equation*}
We can now replace the intractable bound in Eq.[\ref{eq:link_normalizing_Zf}] with its estimate, and obtain an unbiased estimator for the \textit{complete} intractable likelihood of the form as:
\begin{equation*}
    \hat{f}^{(r)}(y_{1:n}|\theta_{1}) = \prod_{i=1}^{n} \frac{q_f(y|\theta)}{\mathcal{Z}_g(\gamma)}\frac{\widehat{M}^{(r)}_{f/g}}{B_{f/g}}.
\end{equation*}
For a more detailed description of the method, refer to Section 3.3 of \citep{Benson21}.

\section{- Full conditionals for the CMP-SAS model}\label{app:full_conditionals}
\begin{itemize}
    \item For the latent assignments \(\bm{Z}\), the posterior factorises as follows:
    \begin{align*}
        P(\bm{Z} &\mid \bm{Y}, \bm{\alpha}, \bm{\beta}, \gamma,\bm{\eta},\bm{p}) \\
        &= \prod_{i = 1}^N P(Z_i \mid \bm{Y}, \bm{\alpha}, \bm{\beta}, \gamma,\bm{\eta},\bm{Z}_{-i},\bm{p}) \\
        &\propto \prod_{i = 1}^N P(\bm{Y}_{i,\cdot} \mid \alpha_i, \bm{\beta}_{-i}, \gamma,  \eta_i, Z_i) \pi(Z_i\mid p_i)\pi(p_i) \\
        & = \prod_{i=1}^N\prod_{j\neq i} \left[f_{CMP}(y_{i,j}^H \mid \mu_{i,j}^H, \nu_i) f_{CMP}(y_{i,j}^A \mid \mu_{i,j}^A, \nu_i)\right] f_{\text{Bern}}(Z_i \mid p_i) f_{\text{Beta}}(p_i \mid a_p,b_p)
    \end{align*}
    where the conditional assignments \(Z_i\) depend only on \(\bm{Y}_{i,\cdot}\), i.e. the observations concerning the scores of team \(i\), and as such is independent of the other assignments \(\bm{Z}_{-i}\) and \(\bm{\eta}_{-i}\). \\
\item For the hierarchical assignment probabilities \(\bm{p}\), the full conditional factorises as
\begin{align*}
P(\bm p \mid \bm Y, \bm \alpha, \bm \beta, \gamma, \bm Z, \bm \eta)
&= \prod_{i=1}^N P(p_i\mid \bm{Y}, \bm{\alpha}, \bm{\beta}, \gamma,\bm{p}_{-i},\bm{Z},\bm{\eta}) = \prod_{i=1}^NP(p_i \mid Z_i).
\end{align*}
Accordingly, the Gibbs update is performed by drawing \(p_i\) independently across \(i\):
\begin{align*}
\bm p \mid \bm Z \;\sim\; \prod_{i=1}^N \text{Beta}\!\left(\alpha_p + Z_i,\; \beta_p + 1 - Z_i\right).
\end{align*}
\item The latent dispersion \(\eta_i\) and attack \(\alpha_i\) parameters are positively correlated and updated jointly, as explained in Section \ref{sec:correlation_att_nu}. Observations of interest are the the scores of team \(i\), \(\bm{Y}_{i \cdot}\). Additionally, the change in \(\alpha_i\) induces a deterministic change in \(\alpha_N\) from the constraint [\ref{eq:stz_constraint}], influencing the terms \(\bm{Y}_{N \cdot}\) of the likelihood. To reflect this change in \(\alpha_N\), \(\eta_N\) is jointly proposed, and the resulting full conditional for \(i = 1,\dots, N-1\) is:
\begin{align*}
\begin{split}
    & P(\alpha_i, \eta_i, \eta_N \mid \bm{Y}, \bm{\alpha}_{-\{i, N\}}, \bm{\beta}, \gamma, \bm{p}, \bm{Z}, \bm{\eta}_{-\{i, N\}}) \\
    &\propto P(\bm{Y}_{i \cdot} \mid \alpha_i, \bm{\beta}_{-i}, \gamma, Z_i, \eta_i)\pi(\alpha_i)\pi(\eta_i) P(\bm{Y}_{N \cdot} \mid\alpha_N, \bm{\beta}_{-N}, \gamma, Z_N, \eta_N)\pi(\eta_N) \\
    &=\prod_{j \neq i} \left[ f_{CMP}(y_{i,j}^H\mid\mu_{i,j}^H,\nu_i) f_{CMP}(y_{i,j}^A\mid\mu_{i,j}^A, \nu_i) \right] f_{\mathcal{N}}(\alpha_i\mid0,\sigma^2_\alpha) f_{\mathcal{N}}(\eta_i\mid0,\sigma^2_\eta) \\
    & \quad \times \prod_{j \neq N} \left[ f_{CMP}(y_{N,j}^H\mid\mu_{N,j}^H,\nu_N) f_{CMP}(y_{N,j}^A\mid\mu_{N,j}^A, \nu_N) \right] f_{\mathcal{N}}(\eta_N\mid0,\sigma^2_\eta),
\end{split}
\end{align*}
which is not tractable given the normalizing constants in \(f_{CMP}\). We employ a MH step with positively correlated bivariate proposal distribution centred at the previous state according to Eq. [\ref{eq:biv_proposal}] to generate \(\{\alpha_i^*, \eta_i^*\}\), and the corresponding induced proposal from Eq.[\ref{eq:fixnu_proposal}] to retrieve \(\{\alpha_N^*, \eta_N^*\}\). Denote the current state space with \(\bm{\theta}\), and the state space with the proposal \(\{\alpha_i^*, \alpha_N^*, \eta_i^*, \eta_N^*\}\) as \(\bm{\theta}^*\). Draw auxiliary data \(\bm{Y_{i, \cdot}'}, \bm{Y}_{N ,\cdot}' \sim \text{CMP}(\bm{\theta}^*)\). The resulting parameters are accepted with probability \(a(\{\alpha_i, \alpha_N, \eta_i, \eta_N\}, \{\alpha_i^*, \alpha_N^*, \eta_i^*, \eta_N^*\})\) equal to:
\par\nobreak\vspace{-0.5cm}{\footnotesize
\begin{align*}
\begin{split}\label{eq:mh_acc_alpha_eta}
A_{\alpha,\eta} = \min \Biggl\{1&, \frac{\prod_{j\neq i} q(y_{i,j}^{H}, y_{i,j}^{A} \mid \bm{\theta}^*)q(y_{i,j}^{H'}, y_{i,j}^{A'} \mid \bm{\theta})}{
\prod_{j\neq i} q(y_{i,j}^{H}, y_{i,j}^{A} \mid \bm{\theta})q(y_{i,j}^{H'}, y_{i,j}^{A'} \mid \bm{\theta}^*)
}\frac{f_{\mathcal{N}}(\alpha_i^* \mid 0, \sigma_\alpha^2)f_{\mathcal{N}}(\eta_i^* \mid 0, \sigma_\eta^2)}{f_{\mathcal{N}}(\alpha_i\mid 0, \sigma_\beta^2)f_{\mathcal{N}}(\eta_i \mid 0, \sigma_\eta^2)}\\
& \times \frac{\prod_{j\neq N} q(y_{N,j}^{H}, y_{N,j}^{A} \mid \bm{\theta}^*)q(y_{N,j}^{H'}, y_{N,j}^{A'} \mid \bm{\theta})}{
\prod_{j\neq N} q(y_{N,j}^{H}, y_{N,j}^{A} \mid \bm{\theta})q(y_{N,j}^{H'}, y_{N,j}^{A'} \mid \bm{\theta}^*)
}\frac{f_{\mathcal{N}}(\eta_N^* \mid 0, \sigma_\eta^2)}{f_{\mathcal{N}}(\eta_N \mid 0, \sigma_\eta^2)}\Biggr\},
\end{split}
\end{align*}
}
where we omit the symmetric proposal kernels \(k(\cdot \mid \cdot) \) and the normalizing constant \(\mathcal{Z(\bm{\theta})}, \mathcal{Z}(\bm{\theta}^*)\) as they cancel out.
\item For the defence parameters \(\beta_i\), the observations of interest will be the goals \textit{conceded} by team \(i\), \( \bm{Y}_{\cdot, i}\). Similarly to the attack case, changes to \(\beta_i\) induce changes to \(\beta_N\). The full conditionals for \(i = 1, \dots, N-1\) is:
\begin{align*}
     P(\beta_i &\mid \ \bm{Y}, \bm{\alpha}, \bm{\beta}_{-i}, \gamma, \bm{\eta}, \bm{Z}, \bm{p}) \\
     & \propto P(\bm{Y}_{\cdot, i} \mid \bm{\alpha}_{-i}, \beta_i, \gamma, \bm{Z}_{-i}, \bm{\eta}_{-i}) \pi (\beta_i) P(\bm{Y}_{\cdot, N} \mid \bm{\alpha}_{-N}, \beta_N, \gamma, \bm{Z}_{-N}, \bm{\eta}_{-N})\\
     &= \prod_{j \neq i} \left[ f_{CMP}(y_{j,i}^H\mid\mu_{j,i}^H,\nu_j) f_{CMP}(y_{j,i}^A\mid\mu_{j,i}^A, \nu_j) \right] f_{\mathcal{N}}(\beta_i\mid0,\sigma^2_\beta) \\
     & \quad \times \prod_{j \neq N} \left[ f_{CMP}(y_{j,N}^H\mid\mu_{j,N}^H,\nu_j) f_{CMP}(y_{j,N}^A\mid\mu_{j,N}^A, \nu_j) \right],
\end{align*}
    which is likewise intractable. We generate a proposal defence parameter \(\beta_i^*\) and the induced change in \(\beta_N^*\) according to Eqs.[\ref{eq:alpha_beta_rwprop}-\ref{eq:stz_constraint}]. Define the current state space with \(\bm{\theta}\), and with \(\bm{\theta}^*\) the state space with the proposal parameters \(\{\beta_i^*, \beta_N^*\}\), and sample auxiliary data \(\bm{Y}_{\cdot, i}', \bm{Y}_{\cdot, N}' \sim \text{CMP}(\bm{\theta}^*)\). The resulting acceptance probability of transitioning from \(\{\beta_i, \beta_N\}\) to \(\{\beta_i^*, \beta_N^*\}\) is:
\par\nobreak\vspace{-0.5cm}{\footnotesize
\begin{align*}
\begin{split}
A_\beta = \min \Biggl\{1&,\frac{ 
\prod_{j\neq i} q(y_{j,i}^{H}, y_{j,i}^{A} \mid \bm{\theta}^*)q(y_{j,i}^{H'}, y_{j,i}^{A'} \mid \bm{\theta})}{
\prod_{j\neq i} q(y_{j,i}^{H}, y_{j,i}^{A} \mid \bm{\theta})q(y_{j,i}^{H'}, y_{j,i}^{A'} \mid \bm{\theta}^*)
}\frac{f_{\mathcal{N}}(\beta_i^* \mid 0, \sigma_\beta^2)}{f_{\mathcal{N}}(\beta_i\mid 0, \sigma_\beta^2)} \\
&\times \frac{
\prod_{j\neq N} q(y_{j,N}^{H}, y_{j,N}^{A} \mid \bm{\theta}^*)q(y_{j,N}^{H'}, y_{j,N}^{A'} \mid \bm{\theta})
}{
\prod_{j\neq N} q(y_{j,N}^{H}, y_{j,N}^{A} \mid \bm{\theta})q(y_{j,N}^{H'}, y_{j,N}^{A'} \mid \bm{\theta}^*)
}\frac{f_{\mathcal{N}}(\beta_N^* \mid 0, \sigma_\beta^2)}{f_{\mathcal{N}}(\beta_N\mid 0, \sigma_\beta^2)} \Biggr\}
\end{split}
\end{align*}
}%
where we omit the symmetric proposal kernels \(k(\cdot \mid \cdot) \) the normalizing constant \(\mathcal{Z(\bm{\theta})}, \mathcal{Z}(\bm{\theta}^*)\) as they cancel out.
\item Lastly, for the home coefficient \(\gamma\), the full conditional is retrieved as:
\begin{align*}
    P(\gamma \mid& \bm{Y}, \bm{\alpha}, \bm{\beta}, \bm{\eta},\bm{Z},\bm{p}) \propto P(\bm{Y}^H \mid \bm{\alpha}, \bm{\beta}, \gamma, \bm{Z}, \bm{\eta}) \pi (\gamma) \\
    &= \prod_{i=1}^N\prod_{j \neq i} \left[ f_{CMP}(y_{i,j}^H\mid\mu_{i,j}^H,\nu_i) \right] \times f_{\mathcal{N}}(\gamma\mid0,\sigma^2_\gamma),
\end{align*}
where we only need to consider the goals scored by the home teams, \(\bm{Y}^H = \{y_{i,j}^H\}_{i = 1, j \neq i}^N\), and is analogously intractable. We proceed in a similar fashion, with a Gaussian proposal:
\begin{align*}
    \gamma^* \sim k(\gamma^* \mid \gamma) =  \text{Normal}(\gamma, s^2_{\gamma}).
\end{align*}
We denote the current state space with \(\bm{\theta}\), and the proposal state space by \(\bm{\theta}^*\). Sample \(\bm{Y}^{H'} \sim \text{CMP}(\bm{\theta}^*)\), and evaluate the transition probability from \(\gamma\) to \(\gamma^*\):
\begin{align*}
    A_\gamma = \min\left\{1, \frac{
    \prod_{i = 1}^N \prod_{j\neq i} q(y_{i,j}^H \mid \bm{\theta}^*) q(y_{i,j}^{H'} \mid \bm{\theta})}{
    \prod_{i = 1}^N \prod_{j\neq i} q(y_{i,j}^H \mid \bm{\theta}) q(y_{i,j}^{H'} \mid \bm{\theta}^*)
    } \frac{f_\mathcal{N}(\gamma^* \mid 0, \sigma^2_\gamma)
    }{f_\mathcal{N}(\gamma \mid 0, \sigma^2_\gamma)
    }\right\}.
\end{align*}
\end{itemize}
\section{Premier League Season 2023/24}\label{app:PL2324}
\begin{table}[htbp]
\centering
\caption{- Summary of 2023/24 Premier League season. Acronyms read \textit{Goals For} (GF), \textit{Goals Against} (GA), \textit{Goals Difference} (GD = GF-GA).}
\label{tab:pl2023_24_summary}
\begin{tabular}{lccccccc}

\textit{Team} & Wins & Draws & Losses & GF & GA & GD & Points \\
\midrule
\textit{Manchester City}   & 28 & 7  & 3  & 96 & 34 & +62 & 91 \\
\textit{Arsenal}           & 28 & 5  & 5  & 91 & 29 & +62 & 89 \\
\textit{Liverpool}         & 24 & 10 & 4  & 86 & 41 & +45 & 82 \\
\textit{Aston Villa}       & 20 & 8  & 10 & 76 & 61 & +15 & 68 \\
\textit{Tottenham} & 20 & 6  & 12 & 74 & 61 & +13 & 66 \\
\textit{Chelsea}           & 18 & 9  & 11 & 77 & 63 & +14 & 63 \\
\textit{Newcastle United}  & 18 & 6  & 14 & 85 & 62 & +23 & 60 \\
\textit{Manchester United} & 18 & 6  & 14 & 57 & 58 & -1  & 60 \\
\textit{West Ham United}   & 14 & 10 & 14 & 60 & 74 & -14 & 52 \\
\textit{Crystal Palace}    & 13 & 10 & 15 & 57 & 58 & -1  & 49 \\
\textit{Brighton} & 12 & 12 & 14 & 55 & 62 & -7 & 48 \\
\textit{Bournemouth}       & 13 & 9  & 16 & 54 & 67 & -13 & 48 \\
\textit{Fulham}            & 13 & 8  & 17 & 55 & 61 & -6  & 47 \\
\textit{Wolves} & 13 & 7 & 18 & 50 & 65 & -15 & 46 \\
\textit{Everton}           & 13 & 9  & 16 & 40 & 51 & -11 & 40 \\
\textit{Brentford}         & 10 & 9  & 19 & 56 & 65 & -9  & 39 \\
\textit{Nott'm Forest} & 9  & 9  & 20 & 49 & 67 & -18 & 32 \\
\textit{Luton Town}        & 6  & 8  & 24 & 52 & 85 & -33 & 26 \\
\textit{Burnley}           & 5  & 9  & 24 & 41 & 78 & -37 & 24 \\
\textit{Sheffield United}  & 3  & 7  & 28 & 35 & 104 & -69 & 16 \\

\end{tabular}
\end{table}

\end{document}